\documentclass[aps,
physrev,
superscriptaddress,
amsmath,
amssymb,
reprint,
nofootinbib,
floatfix]
{revtex4-2}

\usepackage[utf8]{inputenc}
\usepackage{graphicx}
\usepackage{color}
\usepackage{comment}
\usepackage{url}
\usepackage{bm}
\usepackage[italicdiff]{physics}
\usepackage{dcolumn}% Align table columns on decimal point
\usepackage{qcircuit}
\usepackage[version=3]{mhchem}% for molecule names, e.g. \ce{H2O}
\usepackage{siunitx}
\makeatletter
\let\MYcaption\@makecaption
\makeatother

\usepackage{subcaption}
\makeatletter
\let\@makecaption\MYcaption
\makeatother

\newcommand{\qvdots}{% \vdots for qcircuit
  \raisebox{0.3em}{\ensuremath{\vdots}}%
}

\newtheorem{prop}{Proposition}
\DeclareMathOperator{\sign}{sign}

\usepackage{hyperref}
\hypersetup{                    
    colorlinks,
    citecolor=blue,
    filecolor=blue,
    linkcolor=blue,
    urlcolor=blue
}

\begin{document}

\title{
Quantum expectation value estimation\\
by doubling the number of qubits
}

%%% author list
% affiliations
\newcommand{\qunasys}{QunaSys Inc., Aqua Hakusan Building 9F, 1-13-7 Hakusan, Bunkyo, Tokyo 113-0001, Japan}

%authors
\author{Hiroshi Yano}
\affiliation{Department of Applied Physics and Physico-Informatics, Keio University,
Hiyoshi 3-14-1, Kohoku, Yokohama 223-8522, Japan}
\affiliation{\qunasys}

\author{Masaya Kohda}
\author{Shoichiro Tsutsui}
\author{Ryosuke Imai}
\author{\\Keita Kanno}
\affiliation{\qunasys}

\author{Kosuke Mitarai}
\affiliation{Graduate School of Engineering Science, Osaka University, 1-3 Machikaneyama, Toyonaka, Osaka 560-8531, Japan}
\affiliation{Center for Quantum Information and Quantum Biology, Osaka University, Japan}

\author{Yuya O. Nakagawa}
\affiliation{\qunasys}

\date{\today}

\begin{abstract}
    Expectation value estimation is ubiquitous in quantum algorithms. The expectation value of a Hamiltonian, which is essential in various practical applications, is often estimated by measuring a large number of Pauli strings on quantum computers and performing classical post-processing. In the case of $n$-qubit molecular Hamiltonians in quantum chemistry calculations, it is necessary to evaluate $O(n^4)$ Pauli strings, requiring a large number of measurements for accurate estimation. To reduce the measurement cost, we assess an existing idea that uses two copies of an $n$-qubit quantum state of interest and coherently measures them in the Bell basis, which enables the simultaneous estimation of the absolute values of expectation values of all the $n$-qubit Pauli strings. We numerically investigate the efficiency of energy estimation for molecular Hamiltonians of up to 12 qubits. The results show that, 
    when the target precision is no smaller than tens of milli-Hartree,
    this method requires fewer measurements than conventional sampling methods. This suggests that the method may be useful for many applications that rely on expectation value estimation of Hamiltonians and other observables as well when moderate precision is sufficient.
\end{abstract}

\maketitle

%%%%%%%%%%%%%%%%%%%%%%%%%%%%%%%%%%%%%%%%%%%%%%%%%%
\section{Introduction}
Expectation value estimation is a widely performed task in many quantum algorithms, and it plays a crucial role in numerous applications such as quantum chemistry, physical simulations including fluid dynamics, and machine learning. The importance of this task extends not only to hybrid quantum-classical algorithms like variational quantum eigensolver (VQE)~\cite{peruzzo2014Variational}, but also to fault-tolerant quantum computing (FTQC) algorithms (see, e.g., Ref.~\cite{o2022efficient}). Because of its significance, the issue of statistical errors in expectation value estimation has emerged as one of the most prominent challenges toward the practical use of quantum computers.

Typically, the expectation value is estimated by statistically processing the measurement outcomes obtained from quantum computers. To reduce statistical uncertainty, it is necessary to repeat the measurements many times. Specifically, consider the task of estimating the expectation value of an observable $O$, expressed as a linear combination of Pauli strings $P_i \in \{I, X, Y, Z\}^{\otimes n}$ on $n$ qubits. The expectation value can be estimated by aggregating the results of individual projective measurements of each Pauli string $P_i$.
In this scenario,
estimating the expectation value of $O$, with accuracy $\epsilon$ for each Pauli string, requires the total number of measurements to scale as $O(M/\epsilon^2)$, where $M$ is the number of Pauli strings in the observable $O$.

For observables relevant to quantum chemistry and quantum physics, $M$ often grows with the number of qubits $n$. Although the scaling is typically polynomial, it may lead to a practically significant measurement overhead. For instance, in the case of a molecular Hamiltonian with $n$ spin orbitals, the Jordan-Wigner transformation converts the electronic Hamiltonian into a corresponding $n$-qubit Hamiltonian containing $M = O(n^4)$ Pauli strings. Although various techniques have been proposed to reduce this overhead~\cite{mcclean2016Theory,kandala2017Hardwareefficient,izmaylov2019Revising,jena2019Pauli,gokhale2020Measurement,izmaylov2020Unitary,verteletskyi2020Measurement,yen2020Measuring,zhao2020Measurement,hamamura2020Efficient,bonet-monroig2020Nearly,crawford2021Efficient,huggins2021Efficient,yen2021Cartan,shlosberg2023Adaptive,inoue2024Almost,yen2023Deterministic,
huang2020Predicting, % classical shadow 
kohda2022quantum, % computational basis sampling
wecker2015Progress, rubin2018Application}
, they are often insufficient for practical applications~\cite{gonthier2022Measurements,johnson2022Reducing}.

Up to this point, we have implicitly assumed that a single quantum state is prepared and measured at a time. 
However, there is potential for further improvement by preparing multiple copies of a quantum state and performing entangled measurements on them.
One established technique is to prepare two quantum states simultaneously and perform Bell-basis measurements~\cite{huang2021InformationTheoretic, huang2021InformationTheoretic,huang2022Quantum,jiang2020Optimal, garcia2021learning, king2024exponential, chen2024optimal, king2024triply, garcia2021learning}, sometimes referred to as Bell sampling. 
Such a technique has also been widely studied in the context of learning stabilizer states~\cite{montanaro2017Learning}, characterizing stabilizerness and magic~\cite{gross2021Schur,haug2023Scalable,haug2024Efficient}, quantum random sampling~\cite{hangleiter2024Bell}, and eigenstate learning~\cite{grier2024principal}. 
As shown in Ref.~\cite{huang2021InformationTheoretic}, Bell sampling enables one to estimate the absolute value of the expectation value of any Pauli string from the sampling results of a single type of circuit. 
This approach is expected to relax the scaling of the number of shots (i.e., circuit executions) with respect to $M$ for a given accuracy of $\epsilon$. However, it is also known that the shot count can asymptotically scale as $O(1/\epsilon^4)$ in $\epsilon$~\cite{huang2021InformationTheoretic}, which is worse than $O(1/\epsilon^2)$ in the conventional methods by the projective measurements of Pauli strings possibly with some grouping of commuting Pauli strings. 
In addition, a separate estimation of the sign of the expectation value is also required for each $P_i$. Thus, the practical advantage of this method has remained unclear.

In this paper, to clarify the practical advantage of this method, we analyze the bias and the standard deviation in the expectation value estimation of molecular Hamiltonians. In particular, through numerical experiments on some molecular Hamiltonians, we found that, even when accounting for the cost of sign estimation, Bell sampling outperforms the conventional approaches in terms of the shot count for $\epsilon \gtrsim 10~\text{mHa}$. This suggests that Bell sampling is advantageous when performing crude estimates of the expectation values, e.g., during initial iterations in VQE optimization, as proposed in Ref.~\cite{cao2024Accelerated}. Furthermore, as we expect that the cost scaling in $M$ is milder in Bell sampling, it would be possible that, for practical systems where full configuration interaction (Full-CI) calculations are infeasible, Bell sampling maintains advantages even for smaller $\epsilon$.

The structure of this paper is as follows.
Section~\ref{sec:method} provides a brief overview of the method to estimate expectation values of Pauli strings by leveraging the entangled measurements on two copies of a quantum state.
In Sec.~\ref{sec:analy}, we numerically investigate the bias and the standard deviation when we utilize this method for estimating an expectation value of a Hamiltonian expressed by Pauli strings.
In particular, we study molecular Hamiltonians, which are of our interest in quantum chemistry.
Section~\ref{sec:conclusion} concludes the paper with outlook.
Some details are deferred to Appendices: Appendix~\ref{app:eval_method} gives details on the calculation of bias and standard deviation, with an analytic result on the shot-count scaling in the expectation value estimation, while Appendix~\ref{app:other_molecules} provides additional numerical results for dependence on the size of active space for a molecule.

%%%%%%%%%%%%%%%%%%%%%%%%%%%%%%%%%%%%%%%%%%%%%%%%%%
\section{Method}\label{sec:method}

In this section, we describe the method studied in this paper to estimate the expectation value of a Hamiltonian\footnote{
This method can be applied to observables other than Hamiltonians and can be efficient if the observable can be expressed as a linear combination of a polynomial number of Pauli strings $P_i$.
} $H$.
We consider an $n$-qubit quantum state $\ket{\psi}$ to estimate $\ev{H}{\psi}$ for a Hamiltonian expressed as
\begin{equation}\label{eq:n_qubit_hamiltonian}
  H = \sum_{i=1}^M c_i P_i, 
\end{equation}
where
$c_i \in \mathbb{R}$, and $P_i \in \{ I, X, Y, Z \}^{\otimes n}$ for $i = 1, \cdots, M$.
$M$ is the number of Pauli strings $P_i$ contained in $H$.

For this setup, we introduce tensor products of the Pauli strings $P_i \otimes P_i$, dubbed as doubled Pauli strings in this paper. As any pair of doubled Pauli strings commute, all $P_i \otimes P_i$ can be simultaneously measured.
This implies that, for two copies of the quantum state $\ket{\psi} \otimes \ket{\psi}=\ket{\psi} \ket{\psi}$, the expectation values $\bra{\psi}\bra{\psi} P_i \otimes P_i \ket{\psi}\ket{\psi}$ can be simultaneously estimated.
Utilizing $\bra{\psi}\bra{\psi} P_i \otimes P_i \ket{\psi}\ket{\psi} = |\ev{P_i}{\psi}|^2$, one can thus estimate $|\ev{P_i}{\psi}|$ all at once, which are the absolute values of expectation values for Pauli strings in the original system.
In Sec.~\ref{subsec:method_bell}, we present the concrete procedure described in Refs.~\cite{huang2021InformationTheoretic,huang2022Quantum} to realize this idea based on the Bell measurements on the doubled system of $2n$ qubits.
Note that their signs should be estimated separately to obtain the estimate of $\ev{H}{\psi}$. 
Strategies for estimating the signs are discussed in Sec.~\ref{subsec:method_energy}.

\subsection{Measuring Pauli strings in the doubled system}
\label{subsec:method_bell}

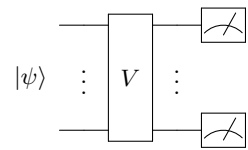
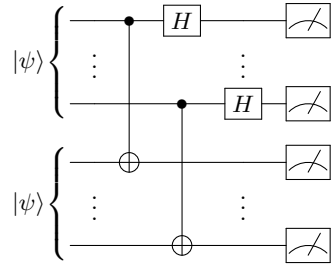
\begin{figure}[tb]
  \begin{subfigure}[t]{0.45\textwidth}
        \mbox{
        \Qcircuit @C=1em @R=1em {
                              & \qw     & \multigate{2}{V} & \qw     & \meter \\
          \lstick{\ket{\psi}} & \qvdots & \nghost{V}       & \qvdots &        \\
                              & \qw     & \ghost{V}        & \qw     & \meter 
        }
      }
      \caption{Conventional sampling}
      \label{fig:standard_qc}
  \end{subfigure}
  \par\bigskip
  \begin{subfigure}[t]{0.45\textwidth}
    \mbox{
      \Qcircuit @C=1em @R=1em {
                              & \qw     & \ctrl{3} & \gate{H} & \qw      & \meter \\
          \lstick{} & \qvdots &          &          & \qvdots  &        \\
                              & \qw     & \qw      & \ctrl{3} & \gate{H} & \meter \\
                              & \qw     & \targ    & \qw      & \qw      & \meter \\
          \lstick{} & \qvdots &          &          & \qvdots  &        \\
                              & \qw     & \qw      & \targ    & \qw      & \meter
          \inputgroupv{1}{3}{1.1em}{1.7em}{\ket{\psi}}
          \inputgroupv{4}{6}{1.1em}{1.7em}{\ket{\psi}}
      }
      }
      \caption{Bell-basis measurements on two copies of a quantum state}
      \label{fig:bell_qc}
  \end{subfigure}
  \caption{Quantum circuits for estimating the expectation value of a Pauli string $P$ for a quantum state $\ket{\psi}$. (a) In the conventional method,
  an additional circuit $V$, depending on the Pauli string, is applied before the measurement in the computational basis. (b) In the method studied in this paper, two copies of $\ket{\psi}$ are prepared and the Bell measurements are performed on each pair of qubits. In this case, the single circuit is enough to measure all the possible Pauli strings, while $V$ has to be tailored for a particular Pauli string in the conventional method.}
  \label{fig:qc}
\end{figure}

We first recapitulate the task of estimating $|\ev{P}{\psi}|$ for all the $n$-qubit Pauli strings $P$,\footnote{
In this subsection, we omit the index of $P_i$ to simplify notation.
}
following Ref.~\cite{huang2021InformationTheoretic}.
They can be estimated through the Bell measurements on $\ket{\psi} \otimes \ket{\psi}$, with the quantum circuit shown in Fig.~\ref{fig:bell_qc}.
For every $k \in \{1,...,n\}$, we jointly measure the $k$-th qubits on the first and second copy of $\ket{\psi}$ in the Bell basis to obtain an measurement outcome $B_k \in \{\Psi^+, \Psi^-, \Phi^+, \Phi^-\}$.
Here, $\ket{B_k}$ stands for either of the Bell states: 
\begin{eqnarray*}
  \ket*{\Psi^+} &= \frac{1}{\sqrt{2}} \left( \ket*{00} + \ket*{11} \right), \\
  \ket*{\Psi^-} &= \frac{1}{\sqrt{2}} \left( \ket*{00} - \ket*{11} \right), \\
  \ket*{\Phi^+} &= \frac{1}{\sqrt{2}} \left( \ket*{01} + \ket*{10} \right), \\
  \ket*{\Phi^-} &= \frac{1}{\sqrt{2}} \left( \ket*{01} - \ket*{10} \right).
\end{eqnarray*}
They are simultaneous eigenstates of $X\otimes X$, $Y\otimes Y$ and $Z\otimes Z$.
Extracting the $k$-th qubit pair from $P\otimes P$ and denoting them as $\sigma_k \otimes \sigma_k$ with $\sigma_k \in \{I, X, Y, Z\}$, $\ket{B_k}$ is an eigenstate of $\sigma_k \otimes \sigma_k$ with an eigenvalue $\lambda_{\sigma_k}(B_k) \in \{-1,+1\}$ for each $k$.
We denote $\bm{B} = (B_1, ..., B_n)$ to summarize the measurement outcomes for $2n$ qubits, or $n$ qubit pairs.
Accordingly, using $\ket{\bm{B}} \equiv \ket{B_1} \otimes \cdots \otimes \ket{B_n}$\footnote{In this definition of $\ket{\bm{B}}$, $2n$ qubits are ordered in a way different from the original ordering in $P\otimes P$ to simplify the expression. It should be understood that the expressions such as $\ev{P \otimes P}{\bm{B}}$ and $\bra{\bm{B}}(\ket{\psi}\ket{\psi})$ are just for notational simplicity and actual calculations are performed with a consistent qubit ordering.}, $|\ev{P}{\psi}|^2$ can be expressed as
\begin{eqnarray*}
    &\mathbb{E}_{\bm{B}} \left[ \ev{P \otimes P}{\bm{B}} \right]
    = \mathbb{E}_{\bm{B}} \left[ \Lambda_P(\bm{B}) \right] \\
    &= \bra{\psi}\bra{\psi} P \otimes P \ket{\psi}\ket{\psi} = |\ev{P}{\psi}|^2,
\end{eqnarray*}
where $\Lambda_{P}(\bm{B}) = \prod_{k=1}^n \lambda_{\sigma_k}(B_k)$ and the expectation $\mathbb{E}_{\bm{B}}$ is taken with respect to the probability mass function $p(\bm{B}) = |\bra{\bm{B}}(\ket{\psi}\ket{\psi})|^2$.

Thus, given measurement outcomes $\{\bm{B}^{(t)}\}_{t=1}^{N_1}$ from $N_1$ repetitions of the Bell measurements, an estimator for $|\ev{P}{\psi}|^2$ can be defined as
\begin{equation}
  \hat{a}(P) = \frac{1}{N_1} \sum_{t=1}^{N_1} \Lambda_{P}(\bm{B}^{(t)}).
\end{equation}
As an estimator for $|\ev{P}{\psi}|$, we adopt the following definition in this paper:
\begin{equation}\label{eq:hat_b}
  \hat{b}(P) = \sqrt{\max(0, \hat{a}(P))},
\end{equation}
where $\max(0,\cdot)$, returning the maximum of $0$ and an input argument, is applied to ensure that the inside of the square root is always non-negative even when $\hat{a}(P)$ is estimated to be negative due to the statistical error.

The definitions of the estimators $\hat{a}(P)$ and $\hat{b}(P)$ follow Ref.~\cite{huang2021InformationTheoretic}, which shows:
\begin{prop}[Corollary 1 of Ref.~\cite{huang2021InformationTheoretic}]\label{thm:pre_single_Pauli}
  Given $N_1 = \Theta(\log(1/\delta)/\epsilon^4)$, for any Pauli string $P$, we have
  \begin{equation}
    \left| \hat{b}(P) - |\ev{P}{\psi}| \right| \leq \epsilon,
  \end{equation}
  with probability at least $1 - \delta$.
\end{prop}
Proposition~\ref{thm:pre_single_Pauli} tells that this method allows us to estimate $|\ev{P}{\psi}|$ for any $P$ simultaneously with high probability, 
but the number of measurements required for the error $\epsilon$ scales as $1/\epsilon^4$, which is worse than the conventional methods with $1/\epsilon^2$.
We stress that the estimator $\hat{b}(P)$ has a non-negligible bias caused by the nonlinearity of the square root and max function in Eq.~\eqref{eq:hat_b}.
In particular, the unfavorable scaling of $1/\epsilon^4$ is due to the max function, which will be discussed in our analysis of Sec.~\ref{sec:analy}.

\subsection{
Estimation of energy expectation value}\label{subsec:method_energy}

We now construct an estimator for the energy expectation value.
In addition to the absolute values $|\ev{P}{\psi}|$, it is necessary to estimate their signs to estimate the expectation value of the Hamiltonian in Eq.~\eqref{eq:n_qubit_hamiltonian}.
Denoting an estimator for the sign of $\ev{P}{\psi}$ by $\hat{s}(P)$, an estimator of $\ev{H}{\psi}$ is given by
\begin{equation}\label{eq:hat_h}
  \hat{h} = \sum_{i=1}^{M} c_i \hat{s}_i \hat{b}_i,
\end{equation}
where $\hat{s}_i \equiv \hat{s}(P_i)$ and $\hat{b}_i \equiv \hat{b}(P_i)$.

The signs may be estimated by the conventional sampling method, potentially equipped with some grouping method. Let $N_2$ be the total of shots to be used for estimating all the signs. 
Assume $[N_2]_{g(i)}$ shots are allocated to a group $g(i)$, which $P_i$ belongs to, and all $[N_2]_{g(i)}$ are odd numbers.
Then, we employ the estimator of the sign as
\begin{align}
    \hat{s}_i
    =
    \begin{cases}
        1 \quad &\left( \sum_{l=1}^{[N_2]_{g(i)}} x_i^{(l)} > 0\right), \\
        -1 \quad &\left( \sum_{l=1}^{[N_2]_{g(i)}} x_i^{(l)} < 0\right),
    \end{cases}
\end{align}
where $x_i^{(l)}=\pm 1$ is the $l$-th measurement outcome of $P_i$.

In Ref.~\cite{huang2021InformationTheoretic}, 
it is also suggested that we only need to estimate signs for any $P$ such that $|\ev{P}{\psi}|$ is relatively large and may set the expectation value of the other $P$ to zero.
This may lead to a reduction in the measurement overhead for the sign estimation.
In Sec.~\ref{ssec:finite_meas}, 
we also examine an option to utilize classical methods to compute the signs of expectation values approximately.

%%%%%%%%%%%%%%%%%%%%%%%%%%%%%%%%%%%%%%%%%%%%%%%%%%
 \section{Performance analysis}\label{sec:analy}
We consider the estimation of the expectation value of a Hamiltonian in Eq.~\eqref{eq:n_qubit_hamiltonian} and investigate the estimation accuracy of the method described in Sec.~\ref{sec:method} in terms of the bias and the standard deviation, given a fixed number of measurements.
We expect that the method may be advantageous with respect to the scaling in the number of Pauli strings $M$ over the conventional methods.
However, there is no guarantee if the method maintains the advantage in the overall estimation of energy expectation value due to the unfavorable scaling $1/\epsilon^4$ in the accuracy $\epsilon$ (see Proposition~\ref{thm:pre_single_Pauli}).
Moreover, $\hat{h}$, the estimator of energy expectation value defined in Eq.~\eqref{eq:hat_h}, can be biased due to, e.g., nonlinear operations of the max and square root in the estimator $\hat{b}(P)$;
the sign estimation can also incur the bias.

Following the standard definitions with some assumptions such as  $\hat{s}_i$ and $\hat{b}_i$ being uncorrelated, the bias and variance of $\hat{h}$ are defined as follows:
\begin{equation}\label{eq:bias_original}
  \mathrm{Bias}[\hat{h}] = \sum_{i=1}^{M} c_i \mathbb{E}[\hat{s}_i] \mathbb{E}[\hat{b}_i] - \ev{H}{\psi},
\end{equation}
\begin{eqnarray}\label{eq:var_original}
  \mathrm{Var}[\hat{h}] &=& \sum_i c_i^2 \mathrm{Var}[\hat{s}_i \hat{b}_i] \nonumber\\
  & &+ \sum_{i \neq j} c_i c_j \mathrm{Cov}[\hat{s}_i \hat{b}_i, \hat{s}_j \hat{b}_j], 
\end{eqnarray}
where
\begin{eqnarray}
  \mathrm{Var}[ \hat{s}_i \hat{b}_i] &=& \mathbb{E}[\hat{b}_i^2] - (\mathbb{E}[\hat{s}_i])^2 (\mathbb{E}[\hat{b}_i])^2, \nonumber\\
  \mathrm{Cov}[\hat{s}_i \hat{b}_i, \hat{s}_j \hat{b}_j] &=& \mathbb{E}[\hat{s}_i \hat{s}_j \hat{b}_i \hat{b}_j] - \mathbb{E}[\hat{s}_i \hat{b}_i] \mathbb{E}[\hat{s}_j \hat{b}_j]. \nonumber
\end{eqnarray}
We used $\mathbb{E}[\hat{s}^2_i] = 1$ for all $i$.
The details of how to evaluate the expectation values appeared above, like $\mathbb{E}[\hat{s}_i]$ and $\mathbb{E}[\hat{b}_i]$, are given in Appendix~\ref{app:eval_method}.
In the next two subsections, we simply assume that the signs are already known 
and numerically investigate the scaling of $\mathrm{Bias}[\hat{h}]$ and the standard deviation $\sqrt{\mathrm{Var}[\hat{h}]}$ with respect to only $N_1$, the number of shots for the Bell measurements.
The effect of the statistical fluctuation and bias caused by the sign estimation with respect to $N_2$ is analyzed in Sec.~\ref{ssec:finite_meas}.

First, the estimation of a single Pauli string is considered in Sec.~\ref{subsec:analy_singlePauli} and we observe that the scaling with respect to $N_1$ agrees with Proposition~\ref{thm:pre_single_Pauli}.
In Sec.~\ref{subsec:analy_molecular}, the estimation of the ground state energy of molecular Hamiltonians is examined.
This is of our interest in that the scaling is considered non-trivial when we make use of the method introduced in Sec.~\ref{sec:method} to estimate the expectation value of a molecular Hamiltonian, which is expressed as a linear combination of Pauli strings.
Finally in Sec.~\ref{ssec:finite_meas}, we consider the sampling cost for sign estimation to conclude the end-to-end cost analysis of the proposed method.

\subsection{Single Pauli string}\label{subsec:analy_singlePauli}
We first evaluate the bias and the standard deviation in the estimation of the absolute value of the expectation value of a single Pauli string, i.e., $H = P$, for different values of $\mu = \bra{\psi} P \ket{\psi}$.
Note that $\ket{\psi}$ is not necessary to be specified as information of the quantum state enters solely via $\mu$ in the calculation of the bias and the standard deviation.
These statistical quantities are calculated in two ways: the summation, as in Eq.~\eqref{eq:exp_b_i}, and the saddle point method, as in Eq.~\eqref{eq:exp_b_i_spm}.
The summation faithfully follows the definition of the bias and the standard deviation, thus producing exact values.
However, as $N_1$ increases, the summation becomes computationally demanding; in such cases, the saddle point method turns out to offer a reasonable approximation.

\begin{figure*}[tb]
  \begin{subfigure}[t]{.45\textwidth}
    \centering
    \includegraphics[width=\linewidth]{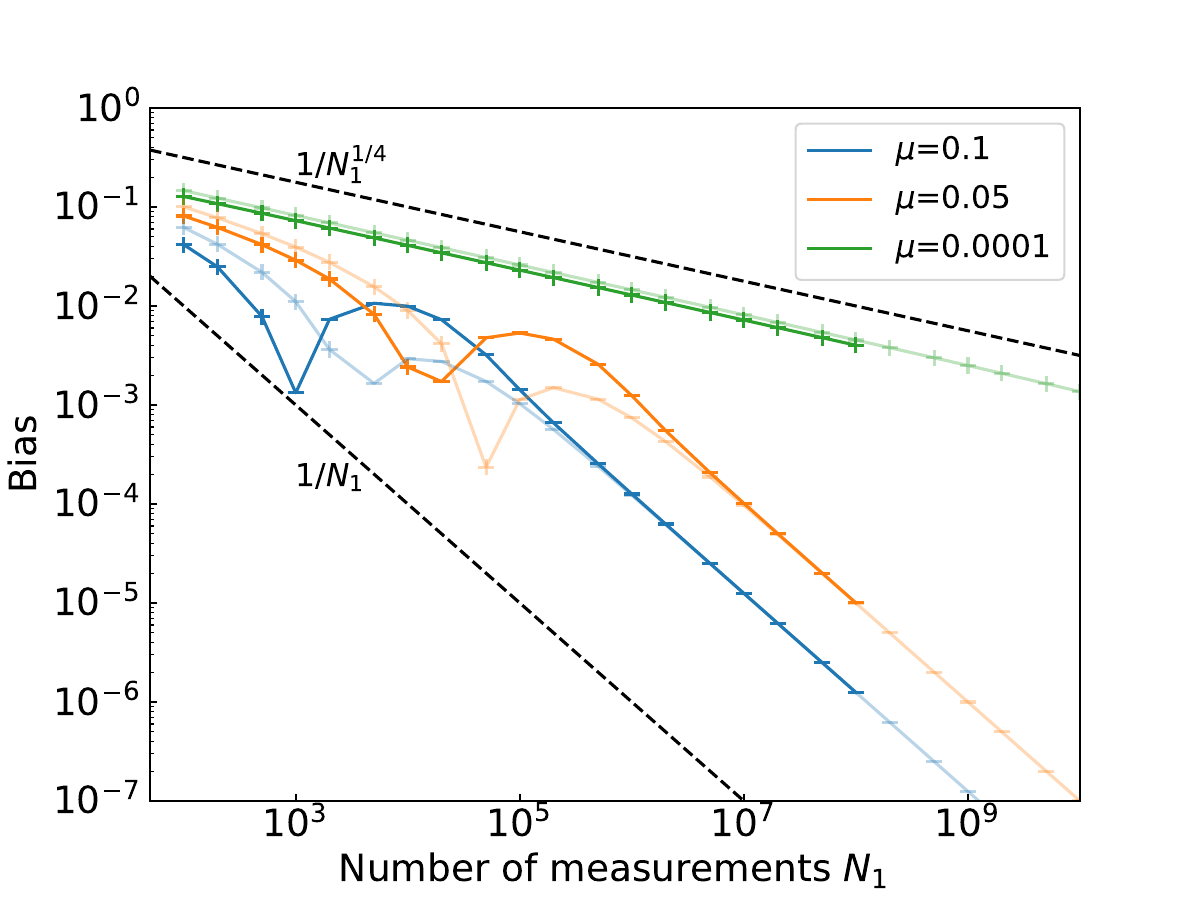}
    \caption{Bias}
    \label{fig:analy_singlePauli_bias}
  \end{subfigure}
  \hfil
  \begin{subfigure}[t]{.45\textwidth}
    \centering
    \includegraphics[width=\linewidth]{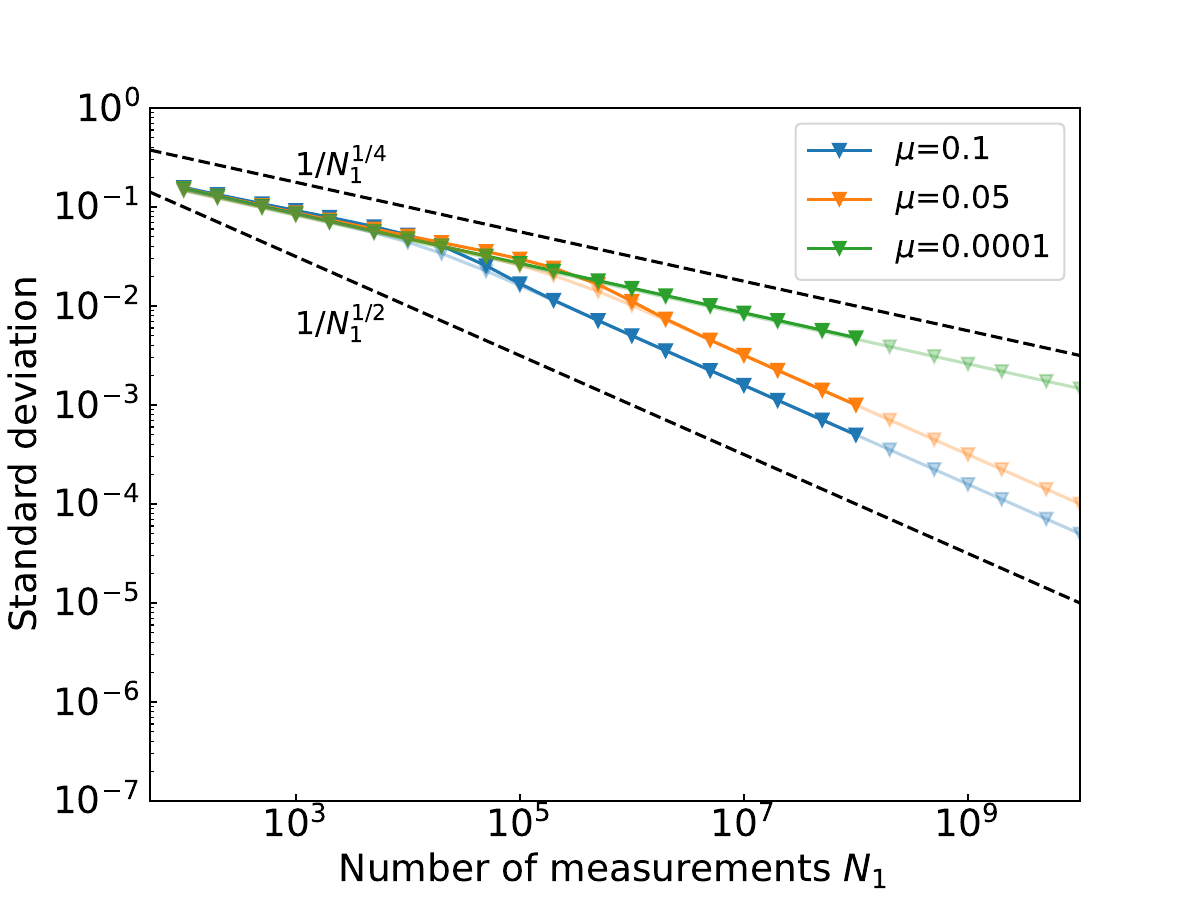}
    \caption{Standard deviation}
    \label{fig:analy_singlePauli_std}
  \end{subfigure}
  \caption{Bias and standard deviation for different values of $\mu = \bra{\psi} P \ket{\psi}$ when estimating the absolute value of the expectation value for a single Pauli string $P$. The symbols '$+$' and '$-$' in (a) represent the sign of bias in Eq.~\eqref{eq:bias_original}. For each $\mu$, the dark line corresponds to the calculation by the summation, as in Eq.~\eqref{eq:exp_b_i}, while the pale line corresponds to the saddle point method, as in Eq.~\eqref{eq:exp_b_i_spm}.}
  \label{fig:analy_singlePauli}
\end{figure*}

The numerical results are shown in Fig.~\ref{fig:analy_singlePauli}.
We see that, in the fewer number of $N_1$, the bias scales as $N_1^{-1}$ for large $\mu$, and $N_1^{-1/4}$ for small $\mu$ close to zero. 
Notably, when increasing $N_1$, the scaling changes from $N_1^{-1/4}$ to $N_1^{-1}$ at some point depending on the value of $\mu$.
The change of the scaling occurs possibly because the fluctuation caused by the sampling error becomes so small that the operation of $\max(0,\cdot)$ is negligible.
However, the bias still remains for larger $N_1$, due to the operation of the square root. 

The standard deviation, on the other hand, scales as $N_1^{-1/2}$ for large $\mu$, and $N_1^{-1/4}$ for small $\mu$ close to zero in the fewer number of $N_1$ (Fig.~\ref{fig:analy_singlePauli_std}). The change in scaling is observed just as in the case of the bias.
We also confirm that the evaluation by the saddle point method (pale line) approximates the one by the summation (dark line) well on the bias for large $N_1$ and the standard deviation. We use this fact to validate the use of the saddle point method for molecular Hamiltonians in the next subsection.

We remark that these observations aligns with Proposition~\ref{thm:pre_single_Pauli}, where the required number of measurements to estimate the absolute value of the expectation value for a single Pauli string with an error $\epsilon$ is given as $O(1/\epsilon^4)$ measurements at most, while we observe a deviation from this upper bound as well.

\subsection{Molecular Hamiltonians without sign estimation}\label{subsec:analy_molecular}

We now study the Hamiltonians for molecules in this subsection. A molecular Hamiltonian is transformed into a linear combination of Pauli strings like Eq.~\eqref{eq:n_qubit_hamiltonian} and the absolute values of expectation values of all the Pauli stings are measured simultaneously by Bell sampling. As shown in the last subsection, the statistical behavior of each Pauli term strongly depends on its expectation value. Those behaviors and the covariance between terms make the scalings of the bias and the standard deviation non-trivial.
Again, we assume that the sign for the expectation value of each Pauli string is given and we will address the problem of sign estimation in the next subsection.

\begin{figure*}[tb]
  \begin{subfigure}[t]{.45\textwidth}
    \centering
    \includegraphics[width=\linewidth]{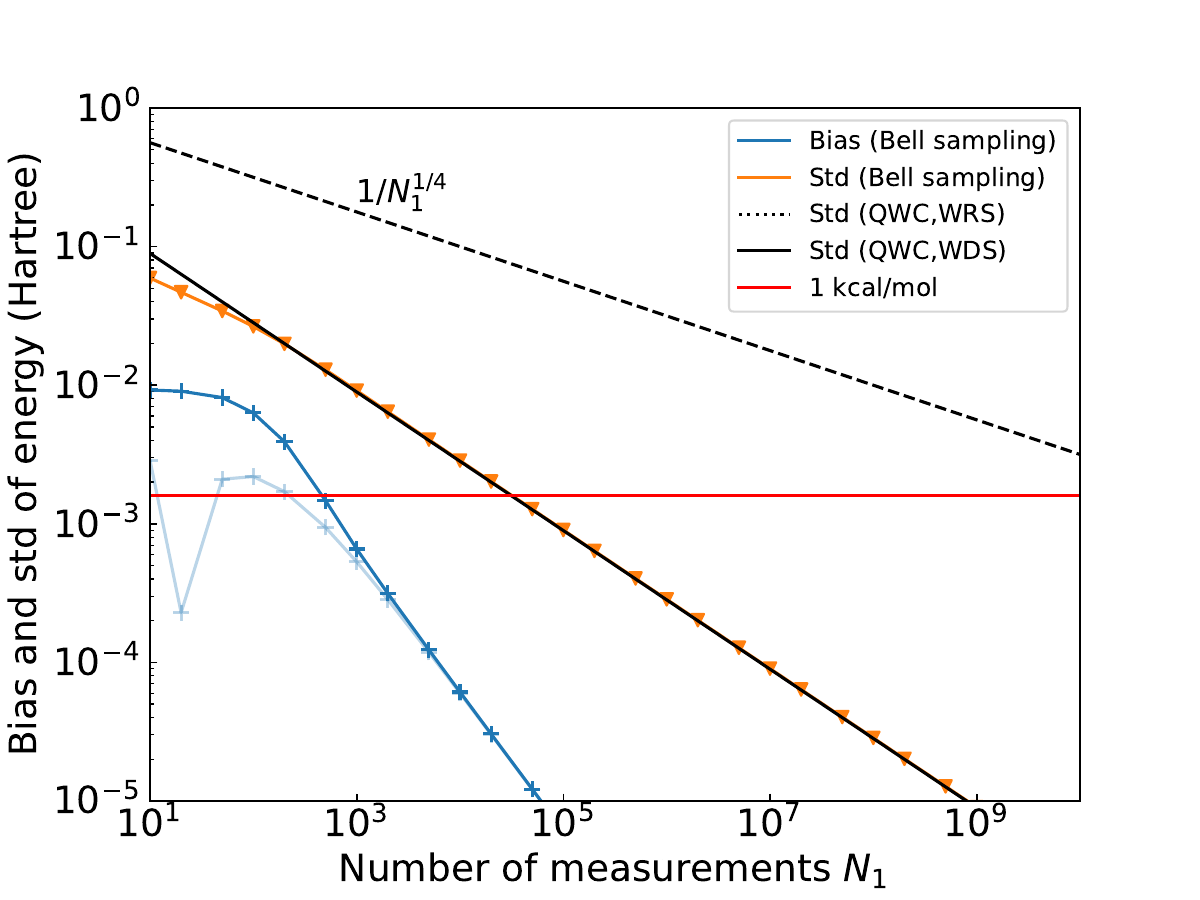}
    \caption{$\mathrm{H_2}$}
    \label{fig:analy_H2}
  \end{subfigure}
  \hfil
  \begin{subfigure}[t]{.45\textwidth}
    \centering
    \includegraphics[width=\linewidth]{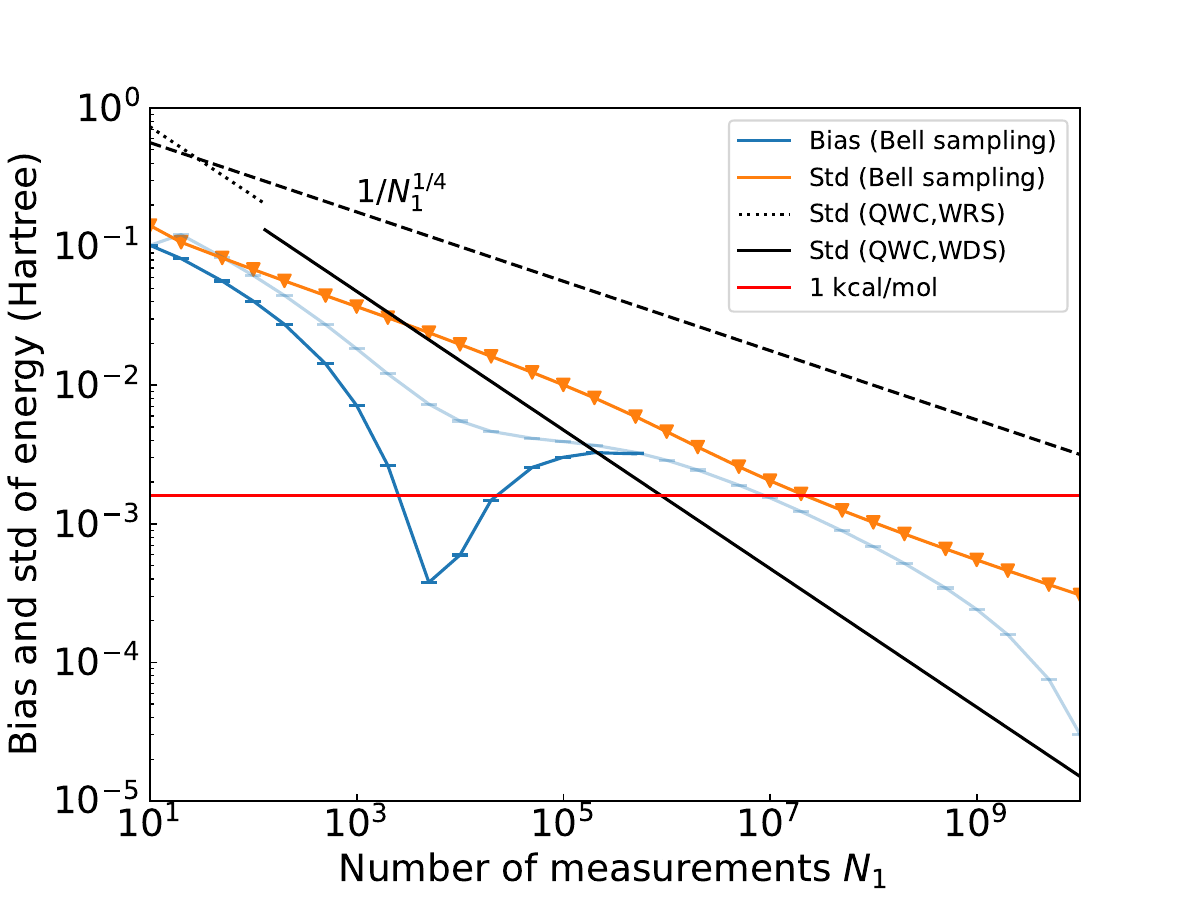}
    \caption{$\mathrm{H_4}$}
    \label{fig:analy_H4}
  \end{subfigure}
  \medskip
  \begin{subfigure}[t]{.45\textwidth}
    \centering
    \includegraphics[width=\linewidth]{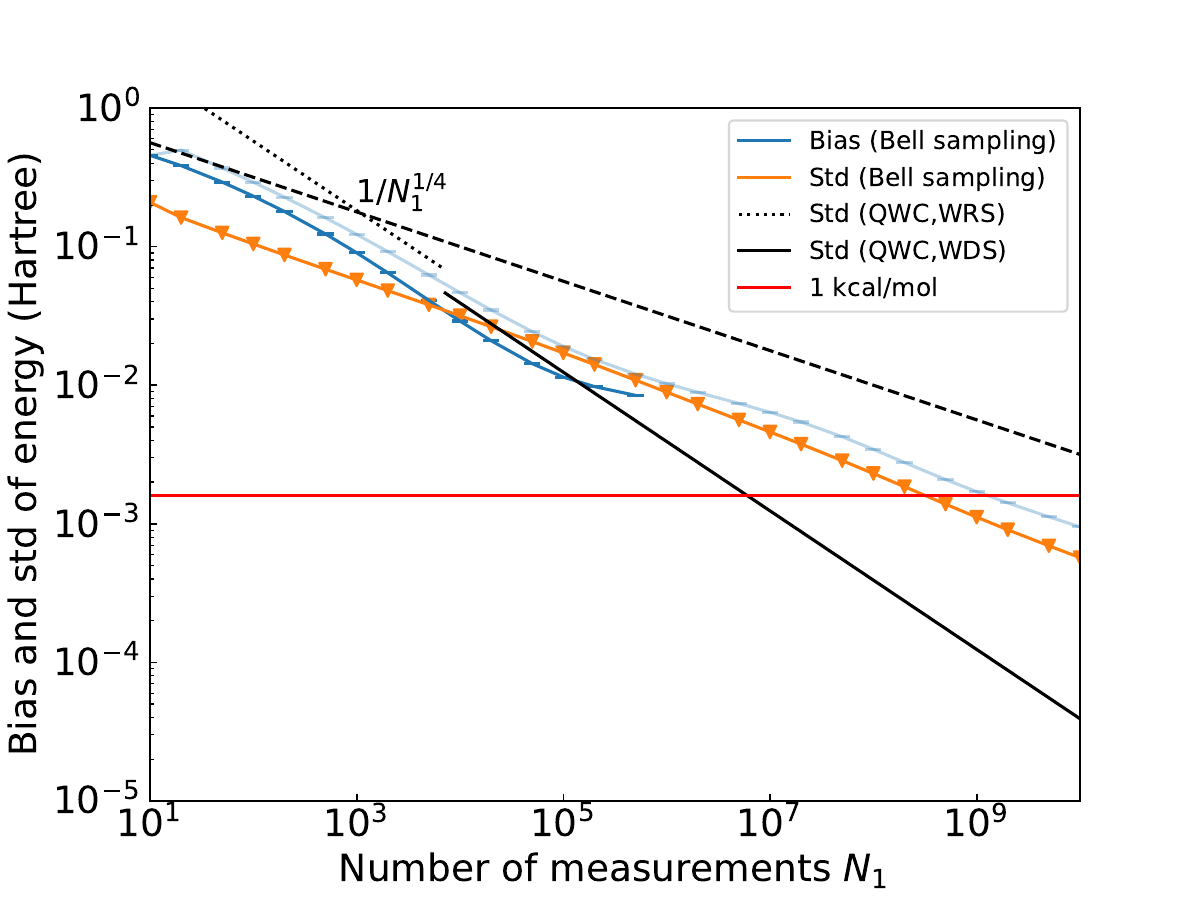}
    \caption{$\mathrm{H_6}$}
    \label{fig:analy_H6}
  \end{subfigure}
  \hfil
  \begin{subfigure}[t]{.45\textwidth}
    \centering
    \includegraphics[width=\linewidth]{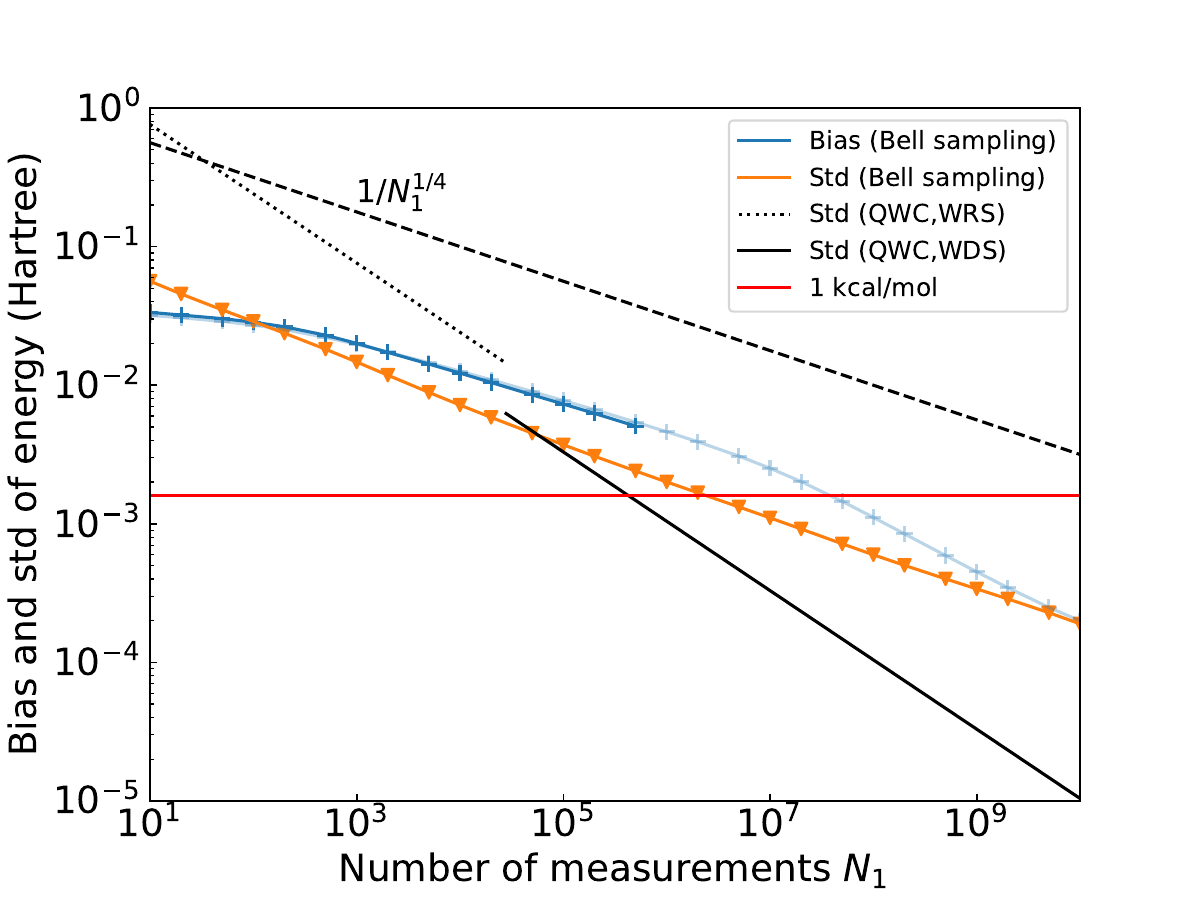}
    \caption{$\mathrm{LiH}$}
    \label{fig:analy_LiH}
  \end{subfigure}
  \caption{Bias and standard deviation on the estimation of the ground-state energy of $\mathrm{H_2}$, $\mathrm{H_4}$, $\mathrm{H_6}$, and $\mathrm{LiH}$ (the sign of the expectation value of each Pauli string in the Hamiltonian is exactly given for the Bell-sampling method in this figure).
  The bias is evaluated using the summation (dark) and the saddle point method (pale), while the standard deviation is evaluated using only the saddle point method.
  The symbols ``$+$'' and ``$-$'' in bias represent its sign.
  The standard deviations of the QWC grouping with WDS/WRS are drawn for comparison in black solid/dotted line, scaling as $N_1^{-1/2}$.
  The red horizontal lines indicate $1$ kcal/mol $\simeq 1.6 \times 10^{-3}$ Hartree. 
  }
  \label{fig:analy_molecules}
\end{figure*}

Molecules considered in our numerical study are hydrogen chains and \ce{LiH}.
We leave the result on \ce{LiH} with active space approximations to Appendix~\ref{app:other_molecules}, where the scaling in the size of the active space is studied.
For the hydrogen chains, we assume each hydrogen atom is equally separated on a line with the distance $1.0$ \AA.
For \ce{LiH}, the atomic distance is $1.6$ \AA.
Those molecular Hamiltonians are considered for electronic states under the Born-Oppenheimer approximation.
The second-quantized fermionic Hamiltonians are produced by OpenFermion~\cite{mcclean2020OpenFermion} interfaced with PySCF~\cite{sun2018PySCF,sun2020Recent}.
The spin orbitals are the Hartree-Fock (HF) orbitals with the STO-3G minimal basis set, and we use the Jordan-Wigner transformation to obtain qubit Hamiltonians from fermionic ones.
Properties of the qubit Hamiltonians are shown in Table~\ref{tab:molecules} for the molecules.
We calculate the exact ground states of the Hamiltonians by the Full-CI method, or exact diagonalization, and take it as the state $\ket{\psi}$ to evaluate the bias and variance of the expectation values of the Hamiltonians (blue and orange lines in Fig.~\ref{fig:analy_molecules}).
To compute expectation values like $\ev{P}{\psi}$, we use numerical libraries Qulacs~\cite{suzuki2021Qulacs} and QURI Parts~\cite{quri_parts}.
As the approximation by the saddle point method is validated in the single Pauli case, the standard deviation is evaluated using only the saddle point method due to computational resource constraint, while the bias is evaluated using both the summation (dark) and the saddle point methods (pale).
\begin{table}[b]
  \caption{\label{tab:molecules}%
  Molecules considered in our numerical study.
  The number of qubits, the number of Pauli strings ($M$) contained in the Hamiltonian (including the identity operator $I$ corresponding to the constant term), and the number of QWC groups are shown for each molecule.
  }
  \begin{ruledtabular}
  \begin{tabular}{cccc}
  \textrm{Molecule}&
  \textrm{Qubits}&
  \textrm{Pauli terms}&
  \textrm{QWC groups}\\
  \colrule
  $\mathrm{H_2}$ & 4 & 15 & 5\\
  $\mathrm{H_4}$ & 8 & 185 & 75\\
  $\mathrm{H_6}$ & 12 & 919 & 329\\
  $\mathrm{LiH}$ & 12 & 631 & 178\\
  \end{tabular}
  \end{ruledtabular}
\end{table}

We compare the accuracy of the estimation against
the conventional method with some grouping strategy, specifically qubit-wise commuting (QWC) grouping~\cite{kandala2017Hardwareefficient,verteletskyi2020Measurement}.
This grouping requires no additional two-qubit gates to simultaneously measure Pauli strings in each group.
Note that there exist more efficient grouping strategies, such as general-commuting grouping~\cite{yen2020Measuring, aaronson2004improved} or almost-optimal groupings whose number of groups are $O(n^2)$~\cite{bonet-monroig2020Nearly,inoue2024Almost}, but they need extra two-qubit gates of more than $O(n)$.
Since Bell sampling only requires $O(n)$ two-qubit gates, QWC grouping is chosen for comparison.
For the shot allocation, i.e. determining the distribution of the number of measurements to each group to be measured, we use two strategies: weighted deterministic sampling (WDS) \cite{wecker2015Progress,rubin2018Application} and weighted random sampling (WRS) \cite{arrasmith2020Operator}, as termed in Ref.~\cite{arrasmith2020Operator}.
In WDS, we assign shots to each group deterministically according to the corresponding weight of each group.
Since WDS cannot be unbiased when the number of shots is less than the number of groups, we also introduce WRS, where shots are assigned to each group randomly according to the weight.
In this paper, the number of measurements that the QWC grouping can use is doubled for a fair comparison with the Bell-sampling method in terms of the number of state preparations of $\ket{\psi}$, i.e., the Bell-sampling method and the QWC grouping are allowed to prepare the same number of copies of $\ket{\psi}$.
Thus, the population standard deviations of the QWC grouping with WDS and WRS are divided by $\sqrt{2 N_1}$ in Fig.~\ref{fig:analy_molecules} (black solid and dotted lines, respectively).

The bias and the standard deviation on the estimation of the ground state energy of \ce{H2} are shown in Fig.~\ref{fig:analy_H2}.
Since the expectation value of each Pauli string in the qubit Hamiltonian of $\mathrm{H_2}$ is larger than $0.1$, the bias and the standard deviation scale almost exactly as $N_1^{-1}$ and $N_1^{-1/2}$, respectively, which reflects the result of the single Pauli case.
We observe that the bias is negligible compared to the standard deviation and the estimation error is comparable with the one of the QWC grouping.

Similar results for 
$\mathrm{H_4}$, $\mathrm{H_6}$, and $\mathrm{LiH}$ are shown in Figs.~\ref{fig:analy_H4}, \ref{fig:analy_H6}, and \ref{fig:analy_LiH}, respectively.
Contrary to the case of $\mathrm{H_2}$, the scaling is non-trivial for both bias and standard deviation.
This is because the expectation values of some of the Pauli strings are small and the changes in their scaling as seen in Fig.~\ref{fig:analy_singlePauli} contribute to the irregular behaviors.
As expected, the efficiency with respect to the number of the Pauli strings by measuring all the Pauli strings at once can be observed when the energy accuracy is around 10-$\SI{30}{mHa}$, compared to the QWC grouping, when the signs are given.
We also reach a similar conclusion for $\mathrm{LiH}$ with various choices of the active space in Appendix~\ref{app:other_molecules}.

 %%%%%%%%%%%%%%%%%%%%%%%%%%%%%%%%%%%%%%%%%%%%%%%%%%
\subsection{Molecular Hamiltonians with sign estimation}\label{ssec:finite_meas}
In this subsection, we finally take into account the sampling cost associated with the sign estimation to assess the end-to-end cost for estimating the expectation value of molecular Hamiltonian.
Here, we propose two directions to handle the signs and investigate the performance in the estimation of ground-state energy for $\mathrm{H_4}$.

First, we numerically study the bias and the standard deviation 
when the signs are estimated by classical computation.
Specifically, we take the CISD calculation, or the configuration interaction (CI) method with single and double excitations above the HF state, for the sign estimation of expectation value of each Pauli string.

\begin{figure}[tb]
  \begin{center}
      \includegraphics[width=.45\textwidth]{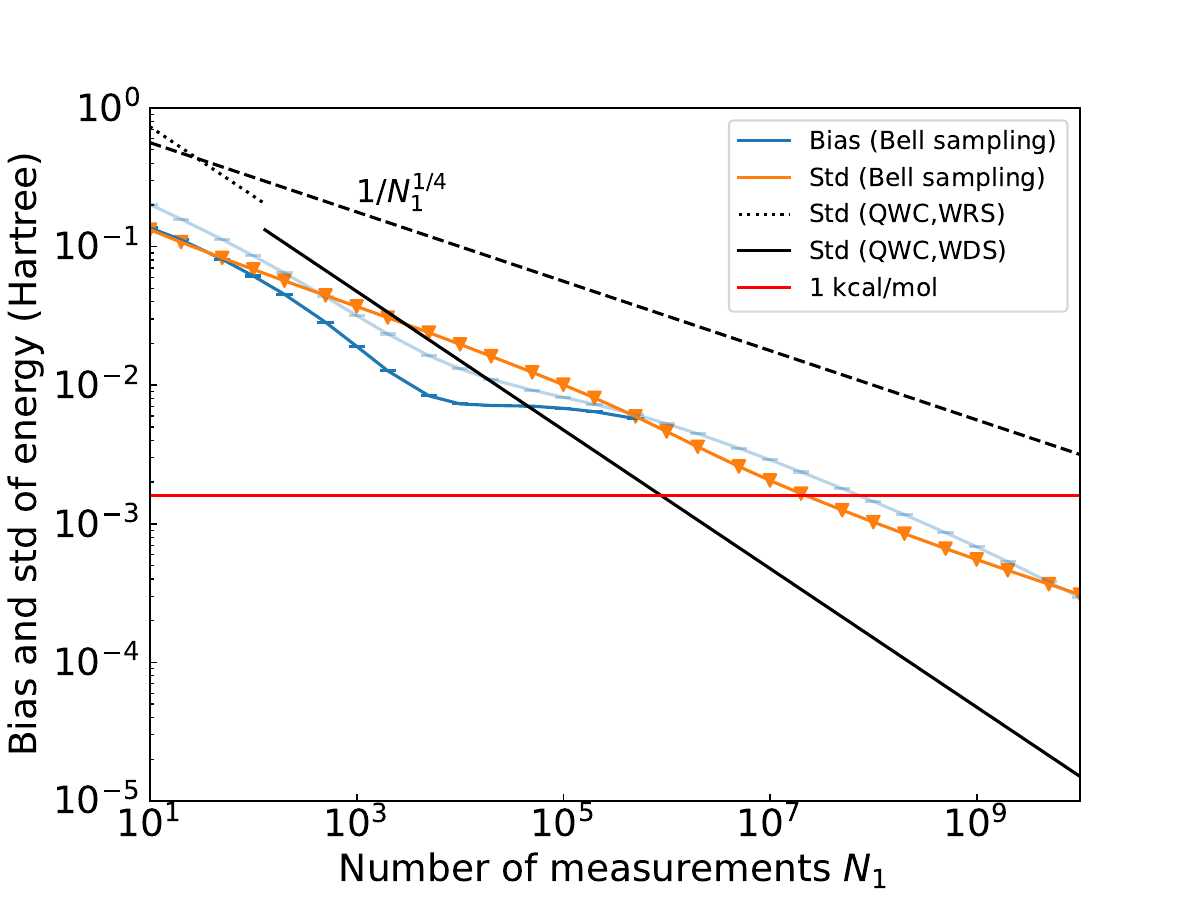}
  \end{center}
  \caption{Bias and standard deviation in the ground-state energy estimation for $\mathrm{H_4}$ with the signs given by the CISD calculation. The results for the conventional QWC grouping are also shown for comparison as before. For details, see the main body in Sec.~\ref{ssec:finite_meas}.}
  \label{fig:analy_H4_cisd}
\end{figure}

Figure~\ref{fig:analy_H4_cisd} shows the result with the signs given by the CISD calculation in the estimation of ground state energy for $\mathrm{H_4}$.
Although the bias gets worse than the case with the correct signs (Fig.~\ref{fig:analy_H4}), there is still an advantage over the QWC grouping in the rough estimation.
This can be attributed to the fact that the truncated CI calculations, such as CISD, can produce a good approximation of the ground-state energy.
However, those truncated CI calculations have computational limitations such as the lack of size consistency, and they may not be able to calculate the signs correctly if, for example, the atomic distance gets larger and the system becomes unstable.
In such cases, it would be worthwhile to consider other classical calculation methods, including coupled cluster methods, for computing the signs of expectation values of Pauli strings.

Next, we investigate a case where the signs are estimated using quantum computers.
For the sign estimation, the conventional sampling method is employed with the QWC grouping, and the shot resources are allocated based on WRS or WDS.
Note that the absolute values, $\hat{b}(P_i)$, are still estimated by Bell sampling. Although conventional sampling is employed, it suffices to estimate the signs of $\ev{P_i}{\psi}$. Hence, the sampling cost regarding this part should be far less demanding than the usual cases where conventional sampling is employed to estimate the entire $\ev{P_i}{\psi}$.

Recall that, given the number of shots $N_1$ and $N_2$, we use $2 N_1$ copies of a quantum state to estimate the absolute values and $N_2$ copies to estimate the signs.
Although the ratio of $N_2$ to $N_1$ should be carefully chosen to reduce the measurement cost, here we show the result for $N_2 = 5 N_1$, which yielded the most favorable outcome among the ratios tested empirically.
To calculate the bias and the standard deviation, we numerically perform sampling experiments, instead of calculating the expectation values shown in Appendix~\ref{app:eval_method}.
In other words, we numerically prepare and measure a quantum state repeatedly to obtain the estimates of the absolute values and the signs in the classical simulation.

\begin{figure}[tb]
  \begin{center}
      \includegraphics[width=.45\textwidth]{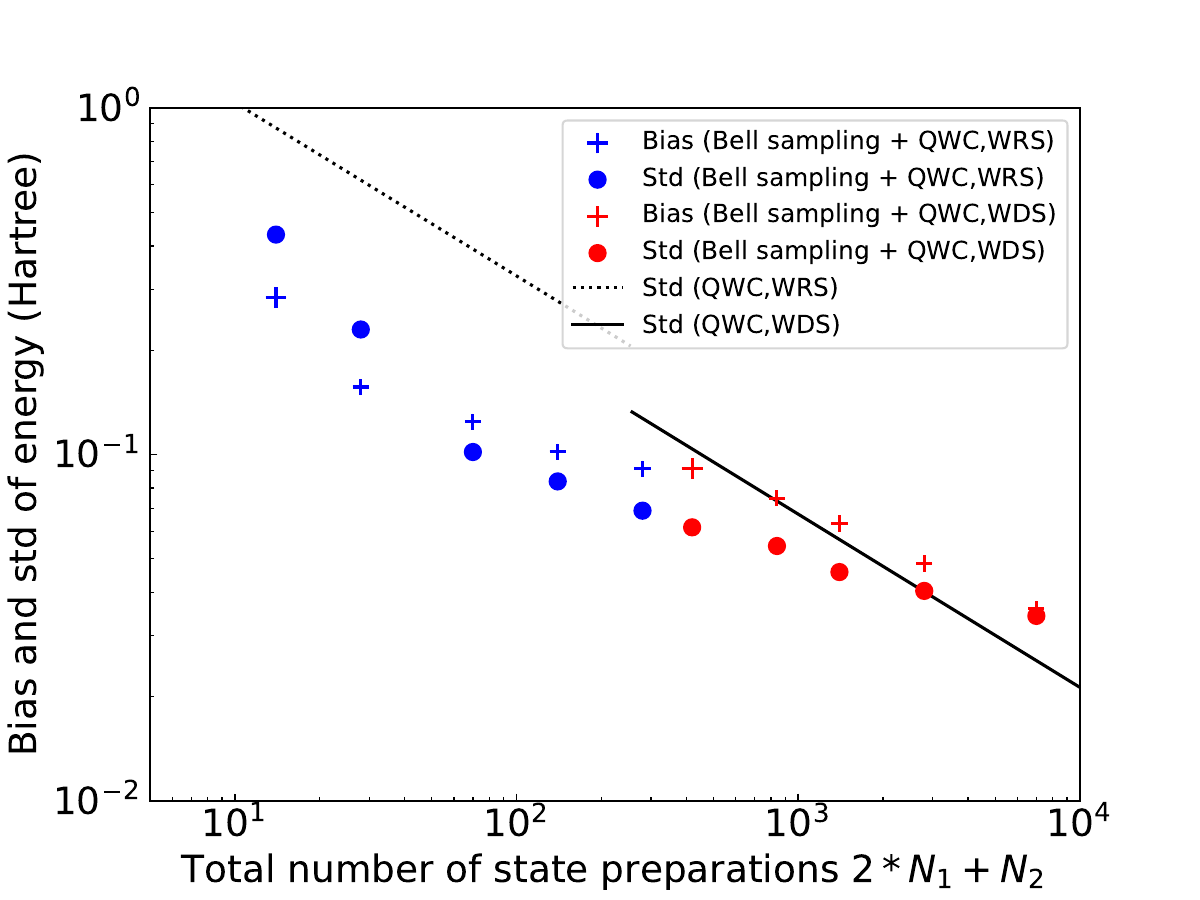}
  \end{center}
  \caption{Bias and standard deviation in the ground-state energy estimation for $\mathrm{H_4}$.
  The signs of $\ev{P_i}{\psi}$ are estimated by the conventional sampling method with $N_2$ shots, while $N_1$ shots for Bell sampling to estimate $\abs{\ev{P_i}{\psi}}$. The results for the conventional QWC grouping are also shown for comparison as before. For details, see the main body in Sec.~\ref{ssec:finite_meas}.
  }
  \label{fig:analy_H4_finiteN2}
\end{figure}

Figure~\ref{fig:analy_H4_finiteN2} shows the bias and the standard deviation in the ground-state energy estimation for $\mathrm{H_4}$ with the signs estimated in the manner described above, based on $1000$ trials.
The Bell-sampling method requires fewer measurements than the conventional method with the QWC grouping for the accuracy up to $10^{-1}$ Hartree, which highlights its possible advantage for the rough estimation, even with the sign estimation.
To suppress the measurement cost of the sign estimation, one could further develop a sampling protocol depending on the estimated absolute values.
It should be a good direction to estimate the signs only for such Pauli strings that their absolute values of expectation values are relatively large, as suggested by Ref.~\cite{huang2021InformationTheoretic}.

%%%%%%%%%%%%%%%%%%%%%%%%%%%%%%%%%%%%%%%%%%%%%%%%%%
\section{Conclusion and outlook}\label{sec:conclusion}
In this study, we assessed the performance of the Bell measurements on two copies of a quantum state, i.e., Bell sampling, in terms of the estimation of expectation values of observables such as Hamiltonians consisting of many Pauli strings.
Bell sampling lets us estimate the absolute values of the expectation values of any Pauli strings at one time, while requiring twice as many qubits as in the conventional methods. 
The expected scaling of Bell sampling and those of conventional methods are different in the required accuracy $\epsilon$ and the number of Pauli strings $M$, so it is a prominent task to identify the parameter region where Bell sampling is advantageous over the conventional sampling methods.
Through the numerical analysis for the estimation of the ground-state energy of molecular Hamiltonians, we identified the situation where this method is advantageous; that is, the rough estimation at the accuracy of several tens of milli-Hartree is more efficient than the conventional sampling methods with the QWC grouping, even when the sign of the expectation value of each Pauli string is not known.
Although it surely depends on the molecular system itself, we expect that this advantage becomes pronounced as the system size gets larger because the scaling advantage in the system size becomes more evident.

When the required accuracy is low, the Bell-sampling method may be more efficient than other conventional estimation methods particularly when there are many Pauli strings to be estimated. The possible application includes the variant of VQE~\cite{zhang2022Variational}, quantum imaginary time evolution~\cite{motta2020Determining}, covariance root-finding~\cite{boyd2022Training}, quantum subspace expansion~\cite{mcclean2017Hybrid,takeshita2020Increasing,mcclean2020Decoding},  algorithmic error mitigation~\cite{suchsland2021Algorithmic}, and shadow spectroscopy~\cite{chan2022algorithmic}.
Moreover, Ref.~\cite{cao2024Accelerated} empirically indicates that the sign estimation of Pauli expectation values can be skipped in the middle of the optimization of VQE.
We expect that our analysis help accelerate the execution of various quantum algorithms through the efficient expectation value estimation.

From a different point of view, instead of using two copies of the quantum state $\ket{\psi} \otimes \ket{\psi}$, we can prepare a different ancilla state $\ket{\chi}$ for the second system and apply the Bell measurements on $\ket{\psi} \otimes \ket{\chi}$.
If the expectation value $\ev{P}{\chi}$ for a Pauli string $P$ is efficiently computable on classical computers, we can estimate $\ev{P}{\psi}$ for every Pauli string with its sign all at once~\cite{jiang2020Optimal}.
The authors of Ref.~\cite{garcia2021learning} propose to determine $\ket{\chi}$ adaptively and show some practical advantage over conventional methods on energy estimation of molecular Hamiltonians.
Reference~\cite{king2024triply} also uses this type of state to show some rigorous efficiency results on estimating expectation values of local fermionic operators and all Pauli operators.
These approaches may point toward promising directions to further enhance the sign estimation.
%%%%%%%%%%%%%%%%%%%%%%%%%%%%%%%%%%%%%%%%%%%%%%%%%%
\section*{Acknowledgments}
We thank T. Yan for suggesting the use of CISD method for sign estimation.
H. Y. is supported by JST SPRING Grant No. JPMJSP2123.
K. M. is supported by JST FOREST Program, Grant Number JPMJFR232Z.

%%%%%%%%%%%%%%%%%%%%%%%%%%%%%%%%%%%%%%%%%%%%%%%%%%
\appendix

%%%%%%%%%%%%%%%%%%%%%%%%%%%%%%%%%%%%%%%%%%%%%%%%%%
\section{Details on evaluating bias and variance}\label{app:eval_method}
In this appendix, we detail how to compute the bias in Eq.~\eqref{eq:bias_original} and the variance in Eq.~\eqref{eq:var_original} for the numbers of measurements $N_1$ and $N_2$.
If we estimate the sign of the expectation value of each Pauli string one by one, the following relations hold:
\begin{eqnarray*}
  \mathbb{E}[\hat{s}_i \hat{s}_j \hat{b}_i \hat{b}_j] &=& \mathbb{E}[\hat{s}_i]\mathbb{E}[\hat{s}_j] \mathbb{E}[\hat{b}_i \hat{b}_j], \\
  \mathbb{E}[\hat{s}_i \hat{s}_j] &=& \mathbb{E}[\hat{s}_i] \mathbb{E}[\hat{s}_j],
\end{eqnarray*}
because the estimation of $\hat{s}_i$ is independent of $\hat{b}_i$ and $\hat{s}_j$ for $i \neq j$. As a result, we need to compute only $\mathbb{E}[\hat{b}_i^2]$ and $\mathbb{E}[\hat{b}_i \hat{b}_j]$, in addition to $\mathbb{E}[\hat{b}_i]$ and $\mathbb{E}[\hat{s}_i]$.
On the other hand, when we measure mutually commuting Pauli strings simultaneously through a grouping strategy, $\hat{s}_i$ is no longer independent of $\hat{s}_j$. Thus, we should compute $\mathbb{E}[\hat{s}_i \hat{s}_j]$ as well.

\subsection{Computation of $\mathbb{E}[\hat{b}_i]$}
Given $N_1$ measurements for measuring an absolute value of the expectation value for any Pauli string $P_i$, the expectation value of each Pauli string is
\begin{equation}\label{eq:exp_b_i}
  \mathbb{E}[\hat{b}_i] = \sum_{m=0}^{N_1} \sqrt{\max\left(0, \frac{2m}{N_1}-1\right)} \binom{N_1}{m} q_i^m (1-q_i)^{N_1-m},
\end{equation}
with the probability of measuring $+1$ on the doubled system
\begin{equation}
  q_i = \frac{1+\bra{\psi}\bra{\psi} P_i \otimes P_i \ket{\psi}\ket{\psi}}{2}.
\end{equation}

When $N_1 \gg 1$, we use the normal approximation of the binomial distribution $\binom{N_1}{m}$ for Eq.~\eqref{eq:exp_b_i} to obtain
\begin{equation}\label{eq:exp_b_i_na}
  \mathbb{E}[\hat{b}_i] \approx \int_{0}^{N_1} dm \sqrt{\max\left(0, \frac{2m}{N_1}-1\right)}f_i^{(\mathrm{normal})}(m),
\end{equation}
where $f_i^{(\mathrm{normal})}(m)$ is the probability density function of the normal distribution in $m$ with mean $N_1 q_i$ and variance $N_1 q_i (1 - q_i)$.

\subsection{Computation of $\mathbb{E}[\hat{b}_i\hat{b}_j]$}
For $i = j$, we can compute $\mathbb{E}[\hat{b}_i^2]$ similarly to $\mathbb{E}[\hat{b}_i]$:
\begin{equation}
  \mathbb{E}[\hat{b}_i^2] = \sum_{m=0}^{N_1} \max\left(0, \frac{2m}{N_1}-1\right) \binom{N_1}{m} q_i^m (1-q_i)^{N_1-m}.
\end{equation}
When $N_1 \gg 1$, according to the normal approximation with $f_i^{(\mathrm{normal})}(m)$ appeared in Eq.~\eqref{eq:exp_b_i_na}, we have
\begin{equation}\label{eq:exp_b_i_squared_na}
  \mathbb{E}[\hat{b}_i^2] \approx \int_{0}^{N_1} dm \max\left(0, \frac{2m}{N_1}-1\right) f_i^{(\mathrm{normal})}(m).
\end{equation}

For $i \neq j$, let $t_i, \, t_j \in \{-1,+1\}$ be the eigenvalues of $P_i \otimes P_i, \, P_j \otimes P_j$, respectively.
Define $m^{(t_i,t_j)}$ to be the number of times we obtain a measurement outcome corresponding to a set of eigenvalues $(t_i,t_j)$.
We then define the multinomial distribution with the probability mass function
\begin{eqnarray}
  g_{ij}^{(\mathrm{multi})}(m^{(+1,+1)},m^{(+1,-1)},m^{(-1,+1)},m^{(-1,-1)}) = \nonumber\\
   N_1! \prod_{t_i,t_j \in \{-1,+1\}} \frac{(q_{ij}^{(t_i,t_j)})^{m^{(t_i,t_j)}}}{m^{(t_i,t_j)}!},
\end{eqnarray}
with the probability of getting a measurement outcome for $(t_i,t_j)$
\begin{equation}
  q_{ij}^{(t_i,t_j)} = \bra{\psi}\bra{\psi} \frac{1+t_i P_i \otimes P_i}{2} \frac{1+t_j P_j \otimes P_j}{2} \ket{\psi}\ket{\psi}.
\end{equation}
For simplicity of notation, the numbers of measurements $m^{(t_i,t_j)}$ are also written as
\begin{eqnarray}\label{eq:m_vec}
  &m_1 = m^{(+1,+1)}, \, m_2 = m^{(+1,-1)}, \, m_3 = m^{(-1,+1)},& \nonumber\\
  &m_4 = m^{(-1,-1)} = N_1 - m_1 - m_2 - m_3.&
\end{eqnarray}
With these notations, we have the expectation value of $\hat{b}_i \hat{b}_j$
\begin{eqnarray}\label{eq:exp_b_i_b_j}
  \mathbb{E}[\hat{b}_i \hat{b}_j] &=& \sum_{m_1=0}^{N_1} \sum_{m_2=0}^{N_1-m_1} \sum_{m_3=0}^{N_1-m_1-m_2} g_{ij}^{(\mathrm{multi})}(m_1, m_2, m_3, m_4) \nonumber\\
  &&\sqrt{\max\left(0, \frac{2(m_1+m_2)}{N_1}-1\right)} \nonumber\\
  &&\sqrt{\max\left(0, \frac{2(m_1+m_3)}{N_1}-1\right)}.
\end{eqnarray}

When $N_1 \gg 1$, we use the normal approximation of the multinomial distribution for Eq.~\eqref{eq:exp_b_i_b_j} to get
\begin{eqnarray}\label{eq:exp_b_i_b_j_na}
  \mathbb{E}[\hat{b}_i \hat{b}_j] &\approx& \int_{m_1=0}^{N_1} \int_{m_2=0}^{N_1-m_1} \int_{m_3=0}^{N_1-m_1-m_2} dm_1 dm_2 dm_3 \nonumber\\
  &&f_{ij}^{(\mathrm{normal})}(m_1, m_2, m_3, m_4) \nonumber\\
  &&\sqrt{\max\left(0, \frac{2(m_1+m_2)}{N_1}-1\right)} \nonumber\\
  &&\sqrt{\max\left(0, \frac{2(m_1+m_3)}{N_1}-1\right)},
\end{eqnarray}
where $f_{ij}^{(\mathrm{normal})}(m_1, m_2, m_3, m_4)$ is the probability density function of the multivariate normal distribution in $m_1, ..., m_4$ with $4$-dimensional mean vector $N_1 \bm q_{ij}$ and $4 \times 4$ covariance matrix $\Sigma_{ij} = N_1(D_{\bm q_{ij}} - \bm q_{ij} \bm q_{ij}^T )$.
Here $\bm q_{ij}$ and $D_{\bm q_{ij}}$ are defined by
\begin{eqnarray*}
  \bm q_{ij} &=& (q_{ij}^{(+1,+1)}, q_{ij}^{(+1,-1)}, q_{ij}^{(-1,+1)}, q_{ij}^{(-1,-1)})^T, \\
  D_{\bm q_{ij}} &=& \mathrm{Diag}(q_{ij}^{(+1,+1)}, q_{ij}^{(+1,-1)}, q_{ij}^{(-1,+1)}, q_{ij}^{(-1,-1)}).
\end{eqnarray*}
If the covariance matrix is not full rank, the multivariate normal distribution does not have a density.
This is the case, for example, when $\bm q_{ij}$ has zero elements.
In such case, we remove all the coordinates of zero elements of $\bm q_{ij}$ so that the multivariate normal distribution redefined with the reduced $\bm q_{ij}$ has a density.

\subsection{Computation of $\mathbb{E}[\hat{s}_i]$}
Given $N_2^{(i)}$ measurements for estimating the sign, the expectation value of $\hat{s_i}$ reads
\begin{equation}\label{eq:exp_s_i}
  \mathbb{E}\left[\hat{s}_i\right] = 1 - 2 \sum_{m=0}^{\frac{N_2^{(i)}-1}{2}} \binom{N_2^{(i)}}{m} p_i^{m} (1-p_i)^{N_2^{(i)}-m},
\end{equation}
with the probability of measuring $+1$ in the standard sampling
\begin{equation}
  p_i = \frac{1+\bra{\psi} P_i \ket{\psi}}{2}.
\end{equation}
Note that $N_2^{(i)}$ is supposed to be an odd number to ensure that the sign is determined.

\subsection{Computation of $\mathbb{E}[\hat{s}_i\hat{s}_j]$}\label{app:eval_method_sign}
For $i = j$, $\mathbb{E}[\hat{s}^2_i] = 1$ trivially holds. Thus, we only consider the case of $i \neq j$.
Let $[N_2]_{g(i)}$ be the number of shots assigned to a group $g(i)$, which $P_i$ belongs to, and $G$ a set of all the groups.
Note that $g(i) = g(j) = g$ for Pauli strings $P_i$ and $P_j$ of the same group, and assume that $\sum_{g \in G} [N_2]_g = N_2$ and $[N_2]_g$ is an odd number. When $i \neq j$ and $P_i$ and $P_j$ belong to different groups, it holds that $\mathbb{E}[\hat{s}_i\hat{s}_j] = \mathbb{E}[\hat{s}_i]\mathbb{E}[\hat{s}_j]$, which reduces to the calculation in the previous section.
If $P_i$ and $P_j$ belong to the same group, 
let $u_i, \, u_j \in \{-1,+1\}$ be the eigenvalues of $P_i, \, P_j$, respectively.
Let $m'^{(u_i,u_j)}$ be the number of times we obtain a measurement outcome corresponding to a set of eigenvalues $(u_i,u_j)$.
$\mathbb{E}[\hat{s}_i \hat{s}_j]$ can be defined with respect to the multinomial distribution, noting that $g=g(i)=g(j)$, 
\begin{align}
  h_{ij}^{(\mathrm{multi})}(m'^{(+1,+1)},m'^{(+1,-1)},m'^{(-1,+1)},m'^{(-1,-1)}) = \nonumber\\
   {[N_2]_{g}}! \prod_{u_i,u_j \in \{-1,+1\}} \frac{(p_{ij}^{(u_i,u_j)})^{m'^{(u_i,u_j)}}}{m'^{(u_i,u_j)}!},
\end{align}
with the probability of getting a measurement outcome for $(u_i,u_j)$
\begin{equation}
  p_{ij}^{(u_i,u_j)} = \bra{\psi} \frac{1+u_i P_i}{2} \frac{1+u_j P_j}{2} \ket{\psi}.
\end{equation}
Then, 
\begin{eqnarray}\label{eq:}
  \mathbb{E}[\hat{s}_i \hat{s}_j] &=& \sum_{m'_1=0}^{[N_2]_{g}} \sum_{m'_2=0}^{[N_2]_{g}-m'_1} \sum_{m'_3=0}^{[N_2]_{g}-m'_1-m'_2} \nonumber\\
  &&h_{ij}^{(\mathrm{multi})}(m'_1, m'_2, m'_3, m'_4) \nonumber\\
  &&\sign\left(\frac{2(m'_1+m'_2)}{[N_2]_{g}}-1\right) \nonumber\\
  &&\sign\left(\frac{2(m'_1+m'_3)}{[N_2]_{g}}-1\right),
\end{eqnarray}
with the notation
\begin{eqnarray}
  &m'_1 = m'^{(+1,+1)}, \, m'_2 = m'^{(+1,-1)}, \, m'_3 = m'^{(-1,+1)},& \nonumber\\
  &m'_4 = m'^{(-1,-1)} = [N_2]_g - m'_1 - m'_2 - m'_3.&
\end{eqnarray}

\subsection{Saddle point method}
Although $\mathbb{E}[\hat{b}_i]$, $\mathbb{E}[\hat{b}^2_i]$ and $\mathbb{E}[\hat{b}_i \hat{b}_j]$ are defined for large $N_1$ in Eqs.~\eqref{eq:exp_b_i_na}, \eqref{eq:exp_b_i_squared_na}, and \eqref{eq:exp_b_i_b_j_na}, respectively, 
it is hard to evaluate the summation numerically.
To circumvent this numerical problem, we introduce the saddle point method.

Define a new variable $x = (2m)/N_1-1$ and Eq.~\eqref{eq:exp_b_i_na} becomes
\begin{eqnarray}\label{eq:exp_b_i_na_x}
  \mathbb{E}[\hat{b}_i] &\approx& \frac{1}{\sqrt{2 \pi \sigma_i^2}} \int_{-1}^{1} dx \sqrt{\max(0, x)} \exp\left( - \frac{(x - \mu_i)^2}{2 \sigma_i^2} \right) \nonumber\\
  &=& \frac{1}{\sqrt{2 \pi \sigma_i^2}} \int_{0}^{1} dx \sqrt{x} \exp\left( - \frac{(x - \mu_i)^2}{2 \sigma_i^2} \right),
\end{eqnarray}
with 
\begin{equation}\label{eq:mu_and_sigma}
  \mu_i = 2q_i - 1, \quad \sigma_i = \sqrt{ \frac{4q_i(1-q_i)}{N_1} }.
\end{equation}
From Eq.~\eqref{eq:exp_b_i_na_x},
\begin{eqnarray}
  \mathbb{E}[\hat{b}_i] &\approx& \frac{1}{\sqrt{2 \pi \sigma_i^2}} \int_{0}^{\infty} dx \sqrt{x} \exp\left( - \frac{(x - \mu_i)^2}{2 \sigma_i^2} \right) \nonumber\\
  &=& \frac{1}{\sqrt{2 \pi \sigma_i^2}} \int_{0}^{\infty} dx \exp\left( - S_i(x) \right),
\end{eqnarray}
where the first approximation is due to $N_1 \gg 1$ and we define
\begin{equation}
  S_i(x) = \frac{(x - \mu_i)^2}{2 \sigma_i^2} - \frac{1}{2} \log x,
\end{equation}
at the last equation.
Then, let $x_i^*$ be the solution to $S_i'(x) = 0, \, x>0$ and we can explicitly write
\begin{equation}
  x_i^* = \frac{\mu_i + \sqrt{\mu_i^2 + 2 \sigma_i^2}}{2}.
\end{equation}
The saddle point method ensures
\begin{equation}\label{eq:exp_b_i_spm}
  \mathbb{E}[\hat{b}_i] \approx \frac{1}{\sqrt{\sigma_i^2 S_i''(x_i^*)}} \exp\left(-S_i(x_i^*)\right).
\end{equation}

$\mathbb{E}[\hat{b}^2_i]$ and $\mathbb{E}[\hat{b}_i \hat{b}_j]$ can be evaluated through the saddle point method similar to $\mathbb{E}[\hat{b}_i]$:
\begin{equation}
    \mathbb{E}[\hat{b}^2_i] \approx \frac{1}{\sqrt{\sigma_i^2 T_i''(x_i^*)}} \exp\left(-T_i(x_i^*)\right),
\end{equation}
where
\begin{equation*}
    T_i(x) = \frac{(x - \mu_i)^2}{2 \sigma_i^2} - \log x,
\end{equation*}
and $x^*_i$ is the solution to $T_i'(x) = 0, \, x>0$.
Using a variable $\bm m = (m_1, m_2, m_3, m_4)^T$ from Eq.~\eqref{eq:m_vec}, instead of $x$,
\begin{equation}
    \mathbb{E}[\hat{b}_i \hat{b}_j] \approx \frac{1}{\sqrt{|\Sigma_{ij}| | U_{ij}''(\bm m_{ij}^*)|}} \exp\left(-U_{ij}(\bm m_{ij}^*)\right),
\end{equation}
where
\begin{eqnarray*}
    U_{ij}(\bm m) &=& \frac{1}{2} (\bm m - N_1 \bm q_{ij})^T \Sigma_{ij}^{-1} (\bm m - N_1 \bm q_{ij}) \\
    && - \frac{1}{2} \log(\frac{2(m_1+m_2)}{N_1}-1) \\
    && - \frac{1}{2} \log(\frac{2(m_1+m_3)}{N_1}-1),
\end{eqnarray*}
and $\bm m_{ij}^*$ is the solution to $U'_{ij}(\bm m) = 0, \bm m > \bm 0$.
In the case of $\mathbb{E}[\hat{b}_i \hat{b}_j]$, however, it is difficult to analytically find saddle points since it involves three variables.
For the numerical evaluation of $\mathbb{E}[\hat{b}_i \hat{b}_j]$, we use \texttt{nsolve} function in SymPy~\cite{Meurer2017SymPy} to numerically find the saddle point $\bm m_{ij}^*$.

\subsection{Best and worst scaling}\label{appendix:scaling}
In Sec.~\ref{subsec:analy_singlePauli}, we numerically observe the scaling of the bias and the standard deviation with respect to $N_1$ for estimating the absolute value of the expectation value of a single Pauli string: the scaling of the bias changes from $O(N_1^{-1/4})$ to $O(N_1^{-1})$ and the standard deviation from $O(N_1^{-1/4})$ to $O(N_1^{-1/2})$.
To analytically confirm these scalings, we discuss two cases: $\mu > O(N_1^{-1/2})$ and $\mu \ll O(N_1^{-1/2})$.
Here we will omit the subscript $i$ and simply write $\mu$ for $\mu_i$, for instance. Throughout the analysis, we assume $\sigma = O(N_1^{-1/2})$.
\vspace{3mm}

$\mu > O(N_1^{-1/2})$: 
The Taylor expansion for $\sqrt{x}$ around $\mu$ in Eq.~\eqref{eq:exp_b_i_na_x} gives
\begin{widetext}
\begin{eqnarray}\label{eq:scaling_exp_large_mu}
  \mathbb{E}[\hat{b}] &\approx& \frac{1}{\sqrt{2 \pi \sigma^2}} \int_{0}^{1} dx \left(  \sqrt{\mu} + \frac{1}{2\sqrt{\mu}}(x-\mu) - \frac{1}{8\mu^{3/2}}(x-\mu)^2 \right) \exp\left( - \frac{(x - \mu)^2}{2 \sigma^2} \right) \nonumber\\
  &\approx& \sqrt{\mu} - \frac{1}{8\mu^{3/2}} \sigma^2,
\end{eqnarray}
\end{widetext}
where we take the Gaussian integral for each term such as
\begin{eqnarray*}
  \int_{0}^{1} dx \exp\left( - \frac{(x - \mu)^2}{2 \sigma^2} \right) &\approx& \int_{-\infty}^{\infty} dx \exp\left( - \frac{(x - \mu)^2}{2 \sigma^2} \right) \\
  &=& \sqrt{2\pi\sigma^2}.
\end{eqnarray*}
The first approximation in the above holds for large $N_1$.
The variance is given by
\begin{equation}
  \mathrm{Var}[\hat{b}] \approx \frac{1}{\sqrt{2 \pi \sigma^2}} \int_{0}^{1} dx \, x \exp\left( - \frac{(x - \mu)^2}{2 \sigma^2} \right) - \mathbb{E}[\hat{b}]^2.
\end{equation}
A similar calculation gives
\begin{equation}\label{eq:scaling_var_large_mu}
  \mathrm{Var}[\hat{b}] \approx \frac{1}{4\mu} \sigma^2 - \frac{1}{64\mu^3} \sigma^4.
\end{equation}
Applying $\sigma = O(N_1^{-1/2})$ to Eqs.~\eqref{eq:scaling_exp_large_mu} and \eqref{eq:scaling_var_large_mu}, we conclude that the scaling of the bias and the standard deviation is $O(N_1^{-1})$ and $O(N_1^{-1/2})$, respectively.
\vspace{3mm}

$\mu \ll O(N_1^{-1/2})$:
Since $\mu \ll \sigma$, Eq.~\eqref{eq:exp_b_i_na_x} is reduced to
\begin{eqnarray}\label{eq:scaling_exp_small_mu}
  \mathbb{E}[\hat{b}] &\approx& \frac{1}{\sqrt{2 \pi \sigma^2}} \int_{0}^{1} dx \sqrt{x} \exp\left( - \frac{x^2}{2 \sigma^2} \right) \nonumber\\
  &=& \frac{1}{\sqrt{2 \pi \sigma^2}} \cdot \frac{\Gamma(3/4,0)-\Gamma(3/4, 1/(2\sigma^2))}{2^{1/4}} \sigma^{3/2} \nonumber\\
  &\to& \frac{\Gamma(3/4,0)}{2^{3/4} \sqrt{\pi}} \sqrt{\sigma},
\end{eqnarray}
as $\sigma \to 0$ (or $N_1 \to \infty$), where $\Gamma(\cdot,\cdot)$ is the incomplete gamma function
\begin{equation*}
  \Gamma(a,z) = \int_{z}^{\infty} dt \, t^{a-1} \exp(-t).
\end{equation*}
The variance is calculated as follows:
\begin{eqnarray}\label{eq:scaling_var_small_mu}
  \mathrm{Var}[\hat{b}] &\approx& \frac{1}{\sqrt{2 \pi \sigma^2}} \int_{0}^{1} dx \, x \exp\left( - \frac{x^2}{2 \sigma^2} \right) - \mathbb{E}[\hat{b}]^2 \nonumber\\
  &\approx& \left( \frac{1}{\sqrt{2\pi}} - \left( \frac{\Gamma(3/4,0)}{2^{3/4} \sqrt{\pi}} \right)^2 \right) \sigma.
\end{eqnarray}
From Eqs.~\eqref{eq:scaling_exp_small_mu} and \eqref{eq:scaling_var_small_mu}, we conclude that the scaling of both bias and the standard deviation is $O(N_1^{-1/4})$.

%%%%%%%%%%%%%%%%%%%%%%%%%%%%%%%%%%%%%%%%%%%%%%%%%%
\section{Performance analysis using active space approximation}\label{app:other_molecules}
\begin{figure*}[tb]
  \begin{subfigure}[t]{.45\textwidth}
    \centering
    \includegraphics[width=\linewidth]{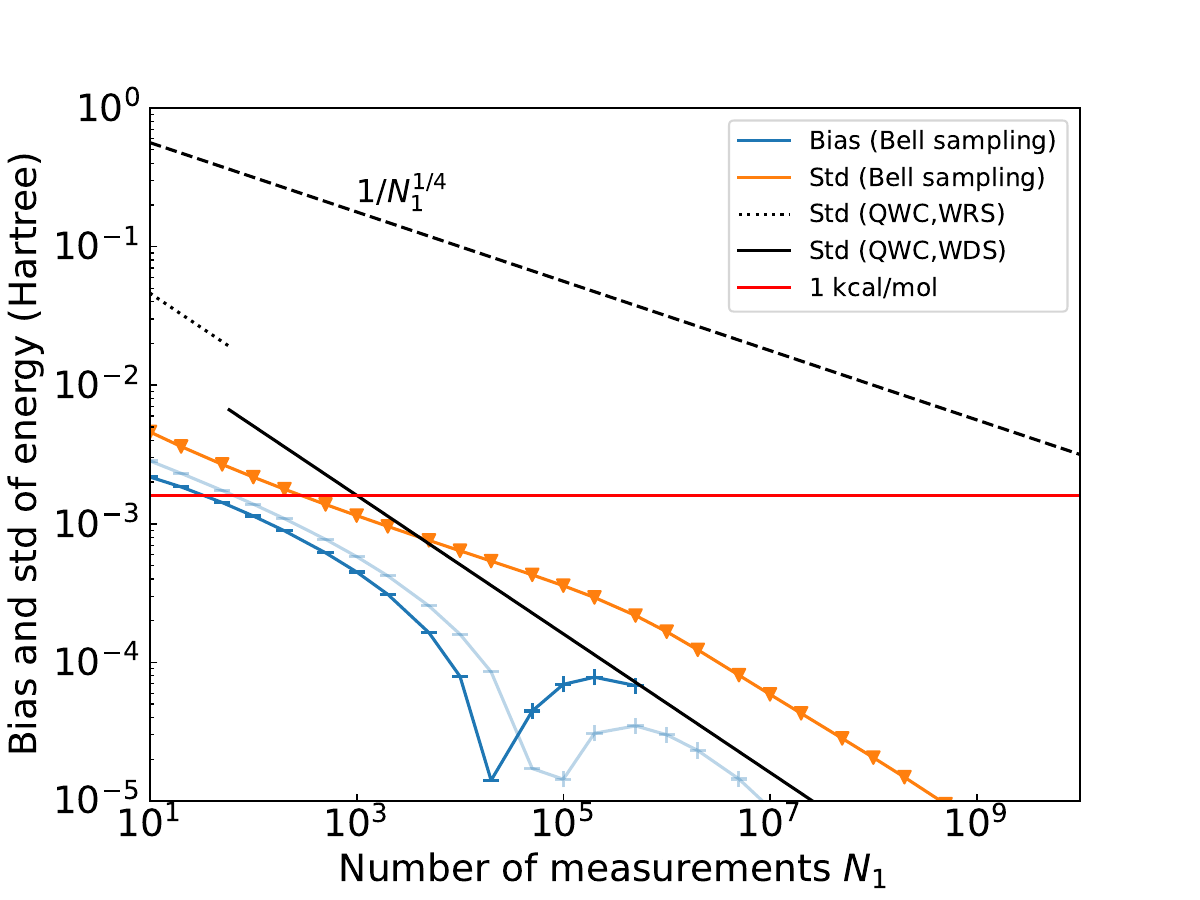}
    \caption{$\mathrm{LiH(2o2e)}$}
    \label{fig:analy_LiH_2o2e}
  \end{subfigure}
  \hfil
  \begin{subfigure}[t]{.45\textwidth}
    \centering
    \includegraphics[width=\linewidth]{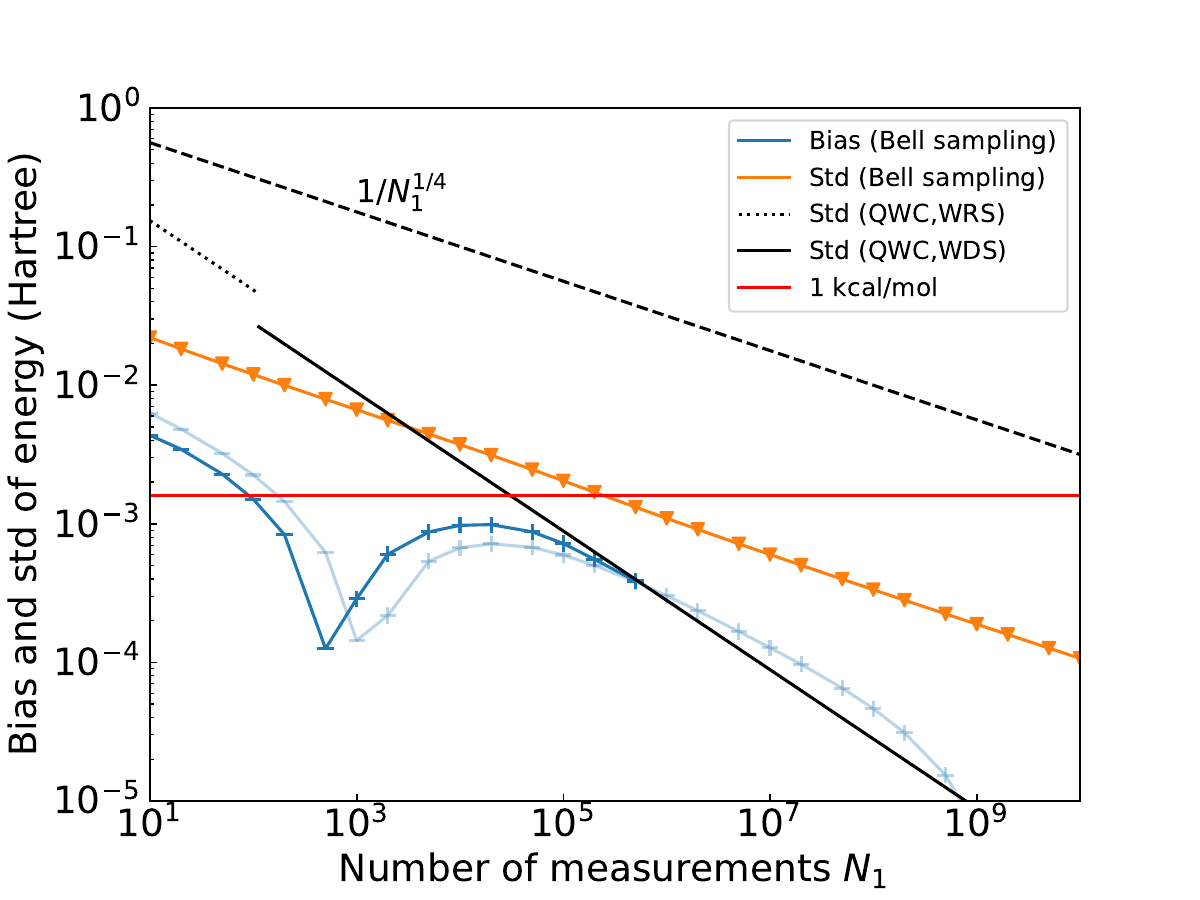}
    \caption{$\mathrm{LiH(4o2e)}$}
    \label{fig:analy_LiH_4o2e}
  \end{subfigure}
  \medskip
  \begin{subfigure}[t]{.45\textwidth}
    \centering
    \includegraphics[width=\linewidth]{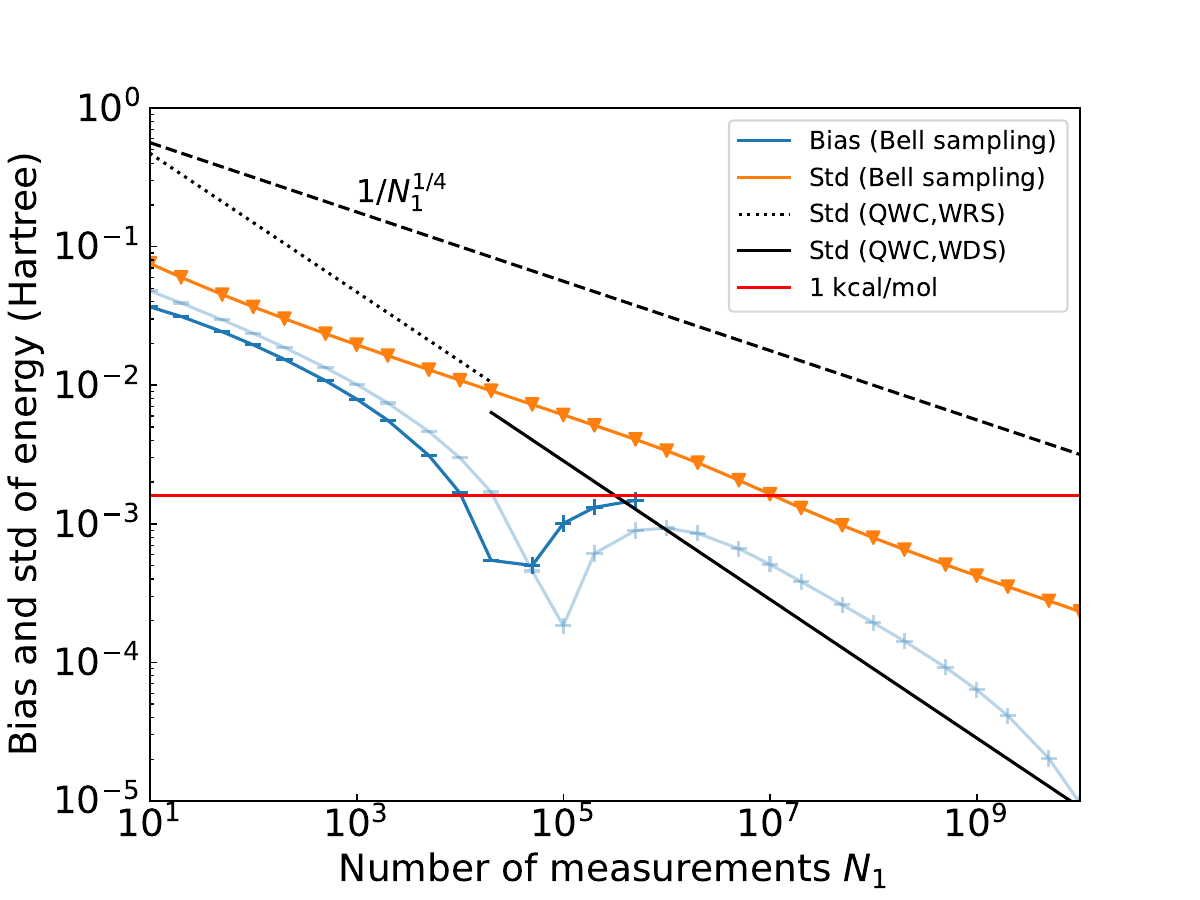}
    \caption{$\mathrm{LiH(4o4e)}$}
    \label{fig:analy_LiH_4o4e}
  \end{subfigure}
  \hfil
  \begin{subfigure}[t]{.45\textwidth}
    \centering
    \includegraphics[width=\linewidth]{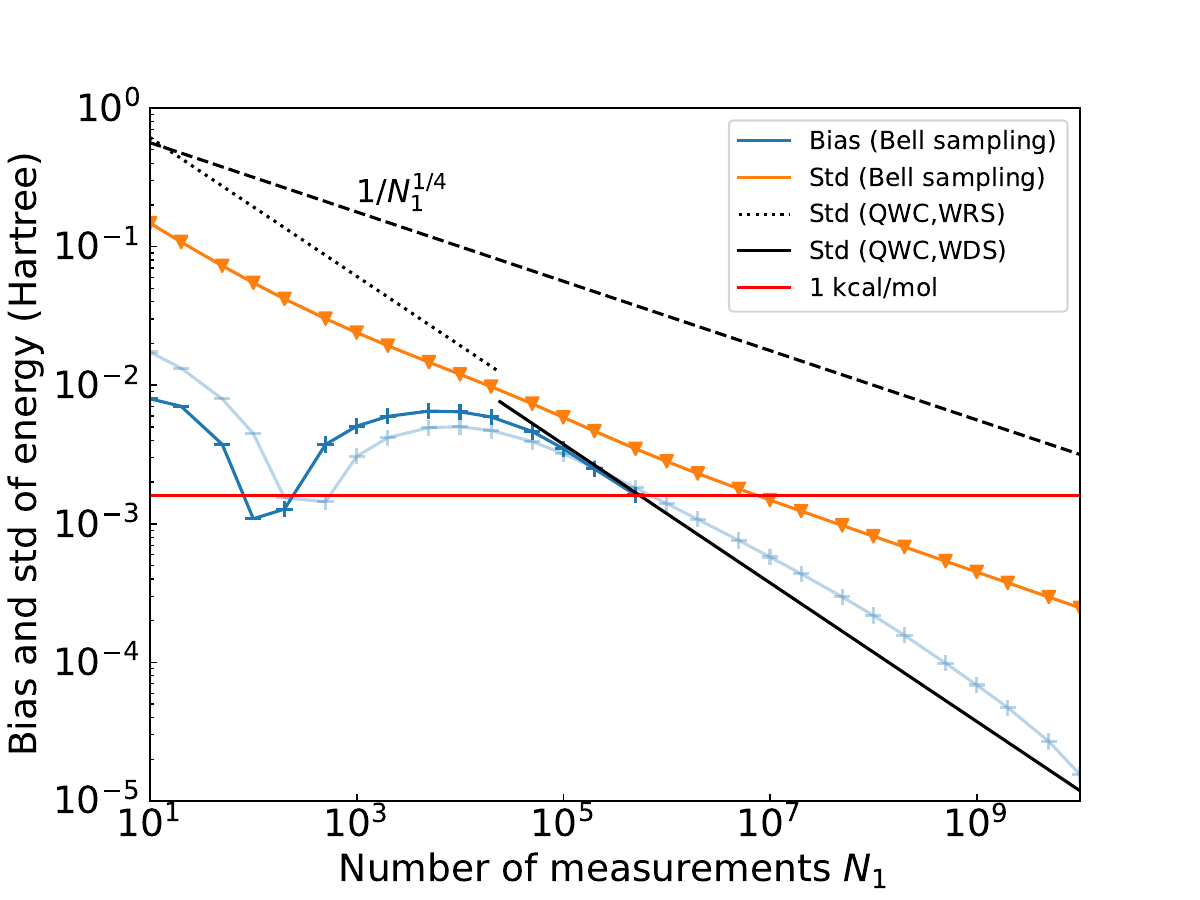}
    \caption{$\mathrm{LiH(5o4e)}$}
    \label{fig:analy_LiH_5o4e}
  \end{subfigure}
  \caption{Bias and standard deviation in the estimation of the ground-state energy of $\mathrm{LiH}$ with the size of active space varied as listed in Table~\ref{tab:molecules_app}. The correct signs are given in the Bell-sampling method. The plots should be interpreted in the same manner as in Fig.~\ref{fig:analy_molecules}.
  }\label{fig:analy_other_molecules}
\end{figure*}

In this appendix, we analyze the bias and standard deviation for $\mathrm{LiH}$ with some patterns of the active space approximation listed in Table~\ref{tab:molecules_app}.
Figure~\ref{fig:analy_other_molecules} shows the bias and standard deviation in the estimation of the ground state energy for each choice of the active space.
Although one might naively expect a smaller bias and standard deviation than the original case of $\mathrm{LiH}$ in Fig.~\ref{fig:analy_LiH} due to the reduction of the system size by the active space approximation, it is not necessarily true for the Bell-sampling method.
For $\mathrm{LiH(2o2e)}$ shown in Fig.~\ref{fig:analy_LiH_2o2e}, we see that the Bell-sampling method is advantageous for even better accuracy, down below the chemical accuracy of $1$ kcal/mol $\simeq 1.6 \times 10^{-3}$ Hartree, in comparison with the QWC grouping.
\begin{table}[b]
  \caption{\label{tab:molecules_app}%
  $\mathrm{LiH}$ molecule of different active spaces considered in this appendix.
  $x$o$y$e stands for the active space with $x$ spatial orbitals and $y$ electrons. $\mathrm{LiH}$ in Table~\ref{tab:molecules} corresponds to the full space of $\mathrm{6o4e}$ in the STO-3G basis set.
  Properties of the qubit Hamiltonians are shown in the same way as Table~\ref{tab:molecules}.
  }
  \begin{ruledtabular}
  \begin{tabular}{cccc}
  \textrm{Molecule}&
  \textrm{Qubits}&
  \textrm{Pauli terms}&
  \textrm{QWC groups}\\
  \colrule
  $\mathrm{LiH(2o2e)}$ & 4 & 27 & 11\\
  $\mathrm{LiH(4o2e)}$ & 8 & 105 & 25\\
  $\mathrm{LiH(4o4e)}$ & 8 & 193 & 45\\
  $\mathrm{LiH(5o4e)}$ & 10 & 276 & 45\\
  \end{tabular}
  \end{ruledtabular}
\end{table}

%%%%%%%%%%%%%%%%%%%%%%%%%%%%%%%%%%%%%%%%%%%%%%%%%%
\bibliography{references.bib}

\end{document}